%% file: main.tex
\documentclass[reprint, superscriptaddress,showpacs,nofootinbib,preprintnumbers,prd]{revtex4-2}

\input{preamble}
\input{acronyms}

\begin{document}

\preprint{FERMILAB-PUB-24-0931-T}

\title{Real-Time Simulation of Asymmetry Generation in Fermion-Bubble Collisions}
\author{Marcela Carena}
\email{mcarena@perimeterinstitute.ca}
\affiliation{Perimeter Institute for Theoretical Physics, 31 Caroline St. N., Waterloo, Ontario N2L 2Y5, Canada}
\affiliation{Fermi National Accelerator Laboratory, Batavia,  Illinois, 60510, USA}
\affiliation{Enrico Fermi Institute, University of Chicago, Chicago, Illinois, 60637, USA}
\affiliation{Kavli Institute for Cosmological Physics, University of Chicago, Chicago, Illinois, 60637, USA}
\affiliation{Department of Physics, University of Chicago, Chicago, Illinois, 60637, USA}
\author{Ying-Ying Li}
\email{liyingying@ihep.ac.cn}
\affiliation{Institute of High Energy Physics, Chinese Academy of Sciences, Beijing 100049, China}
\author{Tong Ou}
\email{tongou@uchicago.edu}
\affiliation{Enrico Fermi Institute, University of Chicago, Chicago, Illinois, 60637, USA}
\affiliation{Department of Physics, University of Chicago, Chicago, Illinois, 60637, USA}
\author{Hersh Singh}
\email{hershs@fnal.gov}
\affiliation{Fermi National Accelerator Laboratory, Batavia,  Illinois, 60510, USA}

\date{\today}

\begin{abstract}
Motivated by the out-of-equilibrium dynamics during an early-universe first-order phase transition, we perform real-time simulations of fermion–bubble scattering in 1+1 dimensions.
This nonequilibrium process can generate a charge-conjugation $\Csym$ asymmetry outside the bubble wall, induced by the complex fermion mass profile.
The resulting $\Csym$ asymmetry is the 1+1-dimensional analog of the $\CP$ asymmetry in 3+1 dimensions, a key ingredient in baryon asymmetry generation at the electroweak scale.
Using tensor network methods, we track the real-time evolution of the $\Csym$ asymmetry in the charge density as the fermion interacts with the bubble wall, a regime inaccessible to analytic calculations.
We further introduce two observables to quantify the asymmetry in the asymptotic region where reflected particles are well separated from the scattering point: one based on the net charge outside the bubble wall, and the other on the spatial displacement between the reflected particle and antiparticle wavepackets.
Our study represents a first step toward nonperturbative, real-time computations of $\CP$ asymmetry in 3+1 dimensions for electroweak baryogenesis.
\end{abstract}

\maketitle

\input{sec_intro}
\input{sec_continuum}
\input{sec_lattice}
\input{sec_conclusion}

\bibliography{ref}

\appendix

\input{app_symmetries.tex}
\input{app_observables}
\input{app_wavepacket}
\input{app_systematics}

\end{document}

%% file: preamble.tex
\usepackage{mathtools}
\usepackage{amsmath,amssymb,amsbsy,amsfonts}

\usepackage{graphics}
\graphicspath{{fig/}{tikz/}}

\usepackage{ninecolors}
\usepackage{soul}
\usepackage[normalem]{ulem}

\usepackage{tikz}
\usetikzlibrary{3d}

\newcommand\Order{O}

\usepackage{siunitx}

\usepackage{microtype}
\usepackage{physics}
\usepackage{physics2}
\usephysicsmodule{ab,ab.braket}

\usepackage{leftindex}

\newcommand\del\partial

\newcommand\Cmat{C}

\usepackage{makecell}
\newcommand\TopRule{\Xhline{0.08em}}

\newcommand\MidRule{\Xhline{0.03em}}
\newcommand\BotRule{\Xhline{0.08em}}

\newcommand\minisec[1]{\textbf{#1}.}

\usepackage{enumitem}
\usepackage{pifont}

\usepackage{graphicx}
\usepackage{amsfonts}
\usepackage{tocvsec2}
\usepackage{slashed}
\usepackage{cancel}
\usepackage{comment}

\usepackage{minibox}
\usepackage{orcidlink}
\usepackage{lipsum}

\newcommand\beq{\begin{eqnarray}}
\newcommand\eeq{\end{eqnarray}}

\newcommand{\mybar}[1]{\kern 0.6pt\overline{\kern -0.6pt#1\kern -0.6pt}\kern 0.6pt}
\usepackage[normalem]{ulem}

\usepackage{xparse}

\newcommand{\psibar}{\Bar{\psi}}

\usepackage[capitalise]{cleveref}
\crefname{section}{Sec.}{Secs.}
\crefname{figure}{Fig.}{Figs.}
\newcommand{\fig}[2]{Fig.~\hyperref[fig:#1]{\ref*{fig:#1}(#2)}}

\usepackage{hyperref}
\hypersetup{
    unicode,
    pdfprintscaling=None,
    colorlinks,
    breaklinks
}
\hypersetup{
  colorlinks=true,   
  allcolors=azure4,
}

\newcommand\Psym{\mathsf{P}}
\newcommand\Csym{\mathsf{C}}

\newcommand\CP{{\Csym\Psym}}

\renewcommand\>{\rangle}

\newcommand\jchi{j_{A}}

\newcommand\jv{j}

\newcommand\Lphys{L}

\newcommand\Lw{L_w}

\newcommand\Qv{Q}

\newcommand\QvsC[2]{\mathcal{Q}_{#2,#1}}
\newcommand\QvsL[2]{\hat{\mathcal{\Qv}}_{#1,#2}}
\newcommand\QvsLf[1]{\hat{\mathcal{\Qv}}_{\infty,#1}}

\newcommand\asymQ[1]{\ensuremath{\mathcal{Q}}_{#1}}
\newcommand\asymQf{\asymQ{\infty}}
\newcommand\asymQL[1]{\ensuremath{\hat{\mathcal{Q}}}_{#1}}
\newcommand\asymQLf{\ensuremath{\hat{\mathcal{Q}}_{\infty}}}

\newcommand\asymP[1]{\ensuremath{\tilde\Phi}(#1)}
\newcommand\asymPf{\ensuremath{\Phi}}
\newcommand\asymPL[1]{\ensuremath{\hat{\Phi}}_{#1}}
\newcommand\asymPLf{\ensuremath{\hat\Phi_{\infty}}}

\newcommand\tauh{\hat{\tau}}

\newcommand{\kbar}{\overline{k_0}}
\newcommand{\sigmabar}{\overline{\sigma_k}}
\newcommand{\sigmaxbar}{\overline{\sigma_x}}
\newcommand{\fbar}{\overline{f}}

\newcommand\ketvac{\ket|\Omega>}

\newcommand\kwp{k_0}
\newcommand\xwp{x_0}

\newcommand\kLat{k}

\newcommand\xwpLat{n_0}
\newcommand\sigmakwp{\sigma_k}
\newcommand\sigmaxwp{\sigma_x}
\newcommand\xwall{x_w}

\newcommand\xdagger{\vphantom{\dagger}}
\newcommand\vacLat{\Omega}
\newcommand\psiLat{\smash{\hat{\psi}}}
\newcommand\Nc{N_c}
\newcommand\jvLat{\smash{\hat\jv}}
\newcommand\tWindowQ{(\Delta t)_{\mathcal{Q}}}
\newcommand\tWindowP{(\Delta t)_{\Phi}}


%% file: acronyms.tex
\usepackage{relsize}

\usepackage[acronyms, nohypertypes={acronym}, nopostdot, style=super, nonumberlist, toc]{glossaries}
\glsenableentrycount
\setacronymstyle{long-sc-short}

\newcommand\ac[1]{\gls{#1}}

\newcommand\acp[1]{\glspl{#1}}

\newacronym{WF}{wf}{Wilson-Fisher}
\newacronym{AF}{af}{asymptotically free}
\newacronym{RG}{rg}{renormalization group}
\newacronym{WZW}{wzw}{Wess-Zumino-Witten}
\newacronym[longplural={conformal field theories}]{CFT}{cft}{conformal field theory}
\newacronym[longplural={lattice field theories}]{LFT}{lft}{lattice field theory}
\newacronym[longplural={effective field theories}]{EFT}{eft}{effective field theory}
\newacronym[longplural={quantum field theories}]{QFT}{qft}{quantum field theory}
\newacronym{LEC}{lec}{low-energy constant}
\newacronym{QCD}{qcd}{quantum chromodynamics}
\newacronym{MC}{mc}{Monte Carlo}
\newacronym{IR}{ir}{infrared}
\newacronym{UV}{uv}{ultraviolet}
\newacronym{SNR}{snr}{signal-to-noise ratio}
\newacronym{NLSM}{nl$\sigma$m}{nonlinear sigma model}
\newacronym{PCM}{pcm}{principal chiral model}
\newacronym{CSA}{csa}{Cartan subalgebra}
\newacronym{SSB}{ssb}{spontaneous symmetry breaking}
\newacronym{DOF}{dof}{degrees of freedom}
\newacronym{DMRG}{dmrg}{density matrix renormalization group}
\newacronym{YM}{ym}{Yang-Mills}
\newacronym{QLM}{qlm}{quantum link model}
\newacronym{KG}{kg}{Kogut-Susskind}
\newacronym{KG-QLM}{kg-qlm}{Kogut-Susskind quantum link model}
\newacronym{D-QLM}{d-qlm}{D-theory quantum link model}
\newacronym{SPT}{spt}{symmetry protected topological} 
\newacronym{GW}{gw}{Ginsparg-Wilson}
\newacronym{FK}{fk}{Fidkowski-Kitaev}
\newacronym{CS}{cs}{Chern-Simons}
\newacronym{APS}{aps}{Atiyah-Patodi-Singer}
\newacronym{PV}{pv}{Pauli-Villars}
\newacronym{PBC}{pbc}{periodic boundary conditions}
\newacronym{OBC}{obc}{open boundary conditions}
\newacronym{ABC}{abc}{antiperiodic boundary conditions}
\newacronym{KD}{kd}{Kahler-Dirac}
\newacronym{RKD}{rkd}{reduced Kahler-Dirac}
\newacronym{SMG}{smg}{symmetric mass generation}

\newacronym{CHN}{chn}{Creutz-Horvath-Neuberger}
\newacronym{GEP}{gep}{generalized eigenvalue problem}
\newacronym{EWBG}{ewbg}{electroweak baryogenesis}
\newacronym{EWPT}{ewpt}{electroweak phase transition}
\newacronym{DWBA}{dwba}{distorted wave Born approximation}
\newacronym{MPS}{mps}{matrix product states}
\newacronym{MPO}{mpo}{matrix product operators}
\newacronym{VIA}{via}{\textsc{vev}-insertion approximation}
\newacronym{FOPT}{fopt}{first-order phase transition}
\newacronym{VEV}{vev}{vacuum expectation values}

%% file: sec_intro.tex
\section{Introduction}

The origin of baryon asymmetry remains one of the most significant questions in particle physics.
Mechanisms for baryogenesis need to satisfy the Sakharov conditions ~\cite{Sakharov:1967dj}: baryon-number violation, $\Csym$ and $\CP$ violation, and departure from thermal equilibrium.
Depending on the temperature of the universe at which baryogenesis occurs, departure from thermal equilibrium is realized by different cosmological processes.
For high-scale baryogenesis, such as \textsc{gut} baryogenesis~\cite{Nanopoulos:1979gx} or leptogenesis~\cite{Yoshimura:1979gy}, the baryon asymmetry is generated at a very high temperature comparable to the Planck mass where the out-of-equilibrium dynamics are due to the expansion of the universe.
For a low-scale baryogenesis, such as \ac{EWBG}~\cite{kuzmin_anomalous_1985,cohen_progress_1993}, the departure from equilibrium is caused by a strong first-order phase transition.

To effectively address out-of-equilibrium dynamics, approximations are applied at multiple stages of the theoretical calculations.
Take the \ac{EWBG} scenario as an example: during the first-order electroweak phase transition, bubbles of true vacuum form, as the Higgs field acquires a \ac{VEV}, and fermions become massive.
In the presence of $\CP$ violation, particles and antiparticles are redistributed among different chiral states in the collisions with the bubble wall, resulting in a nonvanishing chiral asymmetry.
This asymmetry is then converted into net baryon number through the nonperturbative sphaleron process active in the symmetric phase, outside the bubble wall.
The net baryon number generated in front of the bubble wall is subsequently transported into the broken phase as the bubble expands.
If the phase transition is sufficiently strong, the sphalerons become inactive, due to the Boltzmann suppression controlled by the Higgs \ac{VEV}, and the baryon asymmetry is preserved.
Accurate calculations of the bubble wall profile, bubble dynamics, chiral asymmetry generation by particle-bubble collisions, and sphaleron processes present multiple challenges.
These calculations often rely on various approximations, introducing significant theoretical uncertainties to the predictions of the baryon asymmetry \cite{LINDE1980289, cohen_progress_1993, Curtin:2016urg, nucleation, Cline:2020jre,Niemi:2024axp} and other observables, such as gravitational waves emitted from phase transitions \cite{Caprini:2015zlo, Caprini:2019egz, Guo:2020grp, Carena:2022qpf}.

As quantum technology advances, the possibility of achieving fault-tolerant quantum computing systems becomes increasingly attractive and may lead to transformative tools for tackling real-time dynamics in particle physics.
Studies exploring this area are consistently emerging; see review articles \cite{PRXQuantum.4.027001, PRXQuantum.5.037001, Bauer:2023qgm, Fang:2024ple}. Examples include particle scattering problems which are hard to solve by perturbation theory \cite{Jordan:2011ci,chai_fermionic_2024,bennewitz_simulating_2024,farrell_quantum_2024}, parton shower studies considering quantum interference \cite{Bauer:2019qxa, Bepari:2021kwv, Bauer:2023ujy}.
Real-time simulations of out-of-equilibrium dynamics on a quantum computer may well emerge as the best computational tool for a deeper understanding of the evolution of our universe.
Although realistic large-scale quantum simulations are still limited by resources, classical simulations with tensor network methods have enabled real-time studies of scattering phenomenon \cite{PhysRevResearch.3.013078,Rigobello:2021fxw,Papaefstathiou:2024zsu,milsted_collisions_2022,jha_realtime_2024}, as well as dynamics of bubble-wall collisions \cite{milsted_collisions_2022, jha_realtime_2024} in 1 + 1 dimensions.
Tensor network methods exploit the fact that physically relevant states occupy only a small corner of the exponentially large Hilbert space.
In particular, the ground states of the gapped 1+1 dimensional lattice systems can be efficiently approximated by \ac{MPS} \cite{hastings_area_2007,vidal_efficient_2004}.
This insight underlies powerful algorithms such as the \ac{DMRG} \cite{white_density_1992,schollwock_densitymatrix_2005}, which enable efficient computation of ground states and even real-time dynamics in such systems.

In this work, driven by our interest in computing the dynamics of \ac{EWBG}, we present a real-time lattice simulation of particle-bubble scattering using tensor networks.
Our efforts concentrate on a toy model in 1+1-dimensions
to explore the feasibility of real-time simulations to shed light on the realistic 3+1-dimensional scenario.
In particular, we focus on the dynamics of the generation of asymmetry during scattering, commonly calculated with semiclassical methods \cite{Joyce:1994fu,Cline:2000nw}, or \ac{VIA} \cite{Huet:1995mm,Huet:1995sh,Huet:1994jb}.
Both approaches have limitations.
Semiclassical methods only apply to walls with thickness much larger than the de Broglie wavelength of the particle, where the impact of the bubble wall can be approximated by a classical force.
\ac{VIA} evaluates the asymmetry generation with reflection and transmission coefficients, which are only defined in the asymptotic regions away from the scattering point, and thus do not capture the instantaneous asymmetry generated near the scattering point.
Moreover, the calculations of the reflection and transmission coefficients are currently limited to perturbative regimes.
Our study therefore provides the first non-perturbative results on the asymmetry generation over the space throughout the scattering process.
This work marks a necessary step towards simulating baryogenesis from early universe phase transitions on fault-tolerant quantum computers.

This article is organized as follows.
In \cref{sec:con_theory}, we describe our setup for measuring the particle-antiparticle asymmetry in continuum theory.
In \cref{sec:lattice}, we implement this framework on the lattice and present the results of our tensor network simulations.
Finally, in \cref{sec:conclusions}, we discuss the broader implications and outlook of the study.


%% file: sec_continuum.tex
\begin{figure}
    \includegraphics[width=0.85\linewidth]{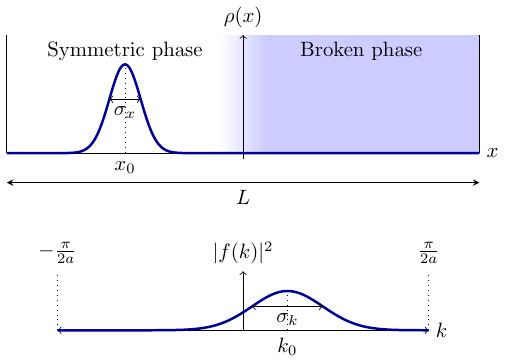}
    \caption{Schematic for our simulations of fermion-bubble scattering. \emph{Top}: Local charge density for a fermion (or anti-fermion) wavepacket in position space. The region around $x=0$ with the color gradient represents the bubble wall. \emph{Bottom:} The wavepacket in momentum space.}
    \label{fig:cartoon}
\end{figure}

\section{Continuum Theory and Observables}
\label{sec:con_theory}

A prototype of chiral asymmetry generation in \ac{EWBG} is a Dirac fermion $\psi$ coupled to a complex scalar field that undergoes a first-order phase transition, from the phase that preserves electroweak symmetry to the one that breaks it.
Bubbles of the true vacuum of the scalar field nucleate and expand, and the massless fermions outside the bubble scatter off the wall.
In this work, we consider the rest frame of the bubble wall, where its effect on the fermions can be described by a complex mass term $m(x)=|m(x)|e^{i\theta(x)\gamma^5}$, where we have $|m(x)|=0$ in the region outside the bubble (symmetric phase $s$), and a nonvanishing mass $|m(x)|\neq 0$ inside the bubble (broken phase $b$). The Hamiltonian in $d+1$ (even) space-time dimensions of the fermion interacting with a static bubble wall is
\begin{align}
    H &= \int \!  d^{d}x\, \bar{\psi} \left[-i\gamma^i\partial_i + |m(x)| e^{i \theta(x) \gamma_{5}}\right] \psi ,
    \label{eq:hcon}
\end{align}
with $i=1, \dotsc, d $.
Dynamical scalar fields can also be incorporated in this framework (see, for example, recent studies of bubble wall collisions \cite{milsted_collisions_2022, jha_realtime_2024}), which we leave for future work.

The presence of a complex mass with $\theta(x) \neq 0$ makes the scattering process qualitatively different from the case of a bubble wall profile with $\theta(x) =0$. This is because a complex mass term generically breaks certain discrete symmetries relating particles and antiparticles.
A nontrivial profile for $\theta(x)$ breaks $\CP$ symmetry in $3+1$-dimensions.
With varying $\theta(x)$, fermions of a certain chirality will scatter differently from their $\CP$ conjugates.

Thus, one can measure changes of the chiral charge density $\jchi^0 = \psibar\gamma^0\gamma^5\psi$ to quantify the effects of $\CP$ violation in scattering.
Since $\jchi^{0}$ is $\CP$ odd, an initially $\CP$ symmetric state would have a vanishing chiral charge density.
With $\CP$-violating interactions, $\jchi^{0}$  would generically become nonzero after scattering.
This net chiral charge density subsequently creates a chemical potential that enables the electroweak sphaleron to generate more baryons than antibaryons.

Can we study this process of $\CP$ asymmetry generation nonperturbatively? While a thorough quantitative study of this problem for the Standard Model is beyond the reach of all current classical or quantum resources, in this work we start an investigation of how we might study this with future quantum hardware.
To do so, we consider the problem of fermion-bubble collisions in a $1+1$-dimensional setting.
In $1+1$-dimensions, the complex-mass term breaks charge-conjugation $\Csym$ (but preserves $\CP$), allowing for chiral asymmetry generation since $\jchi^0$ is odd under $\Csym$.\footnote{For a discussion of the charge-conjugation $\Csym$ and parity $\Psym$ symmetries in arbitrary spacetime dimensions, see \cref{sec:cpt}.}
With the replacement of $\CP$ by charge-conjugation $\Csym$, the discussion of measuring asymmetry generation in the previous paragraph for $(3+1)$-dimensions can be applied to $(1+1)$-dimensions.

The scattering process relevant for generating the asymmetry near the bubble wall is that a Weyl fermion (or its $\Csym$ conjugate anti-fermion) in the symmetric phase moves towards the bubble wall and scatters off.
To develop a formalism for a lattice simulation of this process, let us first describe our setup in the continuum with an infinite volume.

\subsection{Simulation setup in the continuum and infinite volume}

Let $\ketvac$ be the ground state of the Hamiltonian in \cref{eq:hcon}.
We take the static bubble wall profile $m(x)$ to be centered at $\xwall = 0$ with a wall width $\Lw$ such that $m(x) \to 0$ as $x \ll -\Lw$ and $m(x) \to m_0$ for $x \gg \Lw$.
We take the physical volume to be infinite for this discussion.

With $\psi(x),\,{\psi}^{\dagger}(x)$ denoting Dirac fields,
we define the creation operators $\psi^{\dagger}_{k \alpha},$ and the annihilation operators ${\psi}^{\xdagger}_{k \alpha}$ for particles ($\alpha = +$) and antiparticles ($\alpha = -$),
\begin{align}
	\psi(x) = \int \frac{dk}{2\pi}\, \frac{1}{\sqrt{2 \epsilon_k}} \left(  \psi^{\xdagger}_{k+} u_{k} e^{- ikx} + {\psi}^{\dagger}_{k-} v_{k} e^{+ i k x}\right),
    \label{eq:dirac_field_def}
\end{align}
where $u_{k}$ and $v_{k}$ are the positive and negative energy eigenfunctions of the massless Dirac Hamiltonian with energy $\epsilon_k = \pm |k|$.
For a momentum-space wavepacket $f(k)$, we define wavepacket states
\begin{align}
  \ket| f_{\pm} > &= \int dk\, f(k) {\psi}^{\dagger}_{k \pm} \ketvac.
\end{align}
Normalization of the wavepacket states $\braket< f_{\pm} | f_{\pm} > = 1$ requires $\int\! dk\, |f(k)|^2 = 1 / (2\pi)$.
If the wavepacket $f(k)$ is nonvanishing only for $k > 0$, then the states $\ket| f_{\pm} >$ describe a particle or antiparticle wavepacket moving in the positive $x$ direction.
We note that the states $\ket| f_{\pm} >$ are conjugates of each other under charge-conjugation symmetry $\Csym$ that exchange particles and antiparticles:
\begin{align}
\Csym: \psi_{k+}^{\dagger} &\leftrightarrow \psi_{k-}^{\dagger},\\
	\ket| f_{\pm}> &= \Csym \ket| f_{\mp} >.
\end{align}
See \cref{sec:cpt} for a more in-depth discussion of charge conjugation.

As shown in \cref{fig:cartoon},
we take $f(k)$ to be a Gaussian wavepacket localized in momentum space at $k = \kwp > 0$ with width $\sigmakwp$, and localized in position space at $x = \xwp < 0$  with width $\sigmaxwp = 1/(2 \sigmakwp)$,
\begin{align}
f(k; \kwp, \sigmakwp, \xwp) = \frac{1}{(2\pi)^{3/4}\sqrt{\sigma_k}}e^{-ik x_0}e^{-(k-k_0)^2/(4\sigma_k^2)}.
\label{eq:wp-wavefunction}
\end{align}
We need to choose $\sigma_{x}\ll |\xwp|$ to ensure that the wavepacket is located deep inside the massless phase, away from the bubble wall. Furthermore, we need to choose $\sigmakwp \ll \kwp $ to make sure that there are no left-moving modes (modes with negative $k$).
With this,  $\ket|f_{+} >$  and $ \ket| f_{-} >$ are right-moving, left-chiral $(\gamma_5 = -1)$ Weyl fermion and antifermion states, respectively.

Taking the initial state at time $t = 0$ to be either the particle or the antiparticle wavepacket
\begin{align}
  \ket| \Psi_\pm(0)>
  = \ket|f_\pm>,
\end{align}
we turn on time evolution, obtaining the state $\ket| \Psi_\pm(t)> = e^{-i H t} \ket| \Psi_\pm(0)>$ after a time $t$.
As the wavepacket moves towards the bubble wall with velocity $v = 1$, it scatters off the bubble wall at time $t \sim |x_0|$.
Part of the wavepacket is reflected back, whereas the rest is transmitted across the bubble wall into the massive phase.
After a sufficiently long time, the (reflected and transmitted) wavepackets are sufficiently far from the bubble wall.
We can then measure some observables quantifying the asymmetry between how the particle $\ket|f_{+}>$ and antiparticle $\ket|f_{-}>$ wavepackets are scattered.
This would be a measure of the $\Csym$ asymmetry generated during the interactions of the fermions with the bubble wall.

\subsection{Observables}

Breaking of $\Csym$ by the bubble wall implies that the particles scatter differently than the corresponding antiparticles.
Therefore, we need to probe the $\Csym$-asymmetry in the $S$-matrix for the fermion-bubble scattering.

Schematically, the $S$-matrix for elastic fermion-bubble scattering, $S_{\alpha}(k, k')$ ($\alpha=\pm1$ for fermion/anti-fermion), governs how a single-particle momentum eigenstate in the far past $t \to -\infty$,
\begin{align}
  \ket|k \alpha >_{-\infty} = {\psi}^{\dagger}_{k \alpha} \ketvac,
\end{align}
transforms into $\ket|k\alpha>_{+\infty} = \lim_{t \to +\infty} e^{- i H t} \ket| k \alpha >_{- \infty}$ in the far future $t \to + \infty$,
\begin{align}
  \ket|k \alpha>_{-\infty} &\xrightarrow[\text{scattering}]{} \ket| k \alpha>_{+\infty}
  \! = \! \int \! dk' \, S_{\alpha}(k, k') \ket|k'\alpha>_{-\infty}.
\end{align}
Elastic scattering implies that
\begin{align}
  S_{\alpha}(k, k') = R_{\alpha}(k)\delta(k + k') + T_{\alpha}(k) \delta(k' -  k_{T}),
\end{align}
where $k_T = \sqrt{k^{2} - m^2}$ is the transmitted momentum in the massive phase, $R_{\alpha}(k)$ and $T_{\alpha}(k)$ are the complex reflection and transmission coefficients, respectively. Thus,
\begin{align}
\ket| k \alpha>_{+\infty} = R_{\alpha}(k)\ket|-k\alpha>_{-\infty} + T_{\alpha}(k)\ket|k_T\alpha>_{-\infty}
\end{align}
A localized wavepacket $\ket| f_\alpha >_{-\infty} \equiv \int dk f(k)  \ket|k \alpha>_{-\infty}$ in the far past transforms into $\ket| f_\alpha >_{+ \infty}$ in the far future: 
\begin{align}
\ket| f_\alpha >_{+ \infty}
  &= \int\! dk\, f(k) \ab[  R_{\alpha}(k)\ket|-k\alpha>_{-\infty} + T_{\alpha}(k)\ket|k_T\alpha>_{-\infty} ] \notag\\
  &=\ket|f_{R\alpha}>_{+\infty} + \ket|f_{T\alpha}>_{+\infty}
\end{align}
with 
\begin{align}
	\ket|f_{R\alpha}>_{+\infty} 
    &= \int_{}^{}\!\! dk\, R_{\alpha}(k) f(k) \ket| -k\,\alpha >_{-\infty},
                     \label{eq:wavepacket-R}
  \\
	\ket|f_{T\alpha}>_{+\infty} &=  \int_{}^{}\!\! dk\, T_{\alpha}(k)
f(k) \ket|k_T \,\alpha>_{-\infty}.
\end{align}
Let us consider just the reflected wavepacket.
The reflection coefficient $R_{\alpha}(k) = |R_{\alpha}(k)| e^{-i \phi_{\alpha}(k)}$ characterizes the reflected wavepacket completely.
The $\Csym$-asymmetry can appear in both the magnitude $|R_{\alpha}(k)|$ and the phase $\phi_{\alpha}(k)$.
In a lattice simulation, we would like to be able to measure both of these.
Therefore, we define two observables that measure (i) \emph{magnitude asymmetry} and (ii) \emph{phase asymmetry}.

\minisec{Magnitude asymmetry}
Let $\jv^{\mu}(x) = {\bar\psi}(x) \gamma^{\mu} \psi(x)$ be the current for the vector $U(1)$ symmetry, such that the local charge density at time $t$ is
\begin{align}
  \rho_{\pm}(x, t) = \braket< \Psi_\pm(t)| \jv^0(x) | \Psi_\pm(t)>.
  \label{eq:rho-pm}
\end{align}
The total charge in the symmetric phase $x < 0$ is then
\begin{align}
  \QvsC{\pm}{t} &\equiv
  \int_{\mathrlap{-\infty}}^{\mathrlap{0}} dx\, \rho_{\pm}(x,t).
    \label{eq:qvst}
\end{align}
Since the initial wavepacket is completely localized in the symmetric region, $\QvsC{\pm}{0} = \pm 1$ at $t=0$.
At large times, this is related to the reflection magnitudes using \cref{eq:wavepacket-R} and \cref{eq:dirac_field_def}\footnote{This can be seen by writing $\QvsC{\pm}{\infty} = \int_{-\infty}^{0} dx\, \braket< f_{R\pm} | \jv^0(x) | f_{R\pm} >$, noting that the integration limits can be extended to the entire space because $\ket| f_{R\pm} >$ vanishes outside the symmetric region anyway.}:
\begin{align}
\QvsC{\pm}{\infty}
  &= \int_{\mathrlap{-\infty}}^{\mathrlap\infty}\,dx\, \braket< f_{R\pm} | \jv^{0}(x) |f_{R\pm} >_{+\infty} \\
&=  \pm \int_{\mathrlap{-\infty}}^{\mathrlap{\infty}}\, dk\, \ab| R_{\pm}(k) |^{2} \ab|f(k)|^{2}.\label{eq:Qs_inf}
\end{align}
Therefore, a convenient measure of the magnitude asymmetry is simply the sum over the particle and antiparticle states $\asymQ{t} = \mathcal{Q}_{t, +}+ \mathcal{Q}_{t, -}$, such that in the limit $t \to \infty$, we get
\begin{align}
  \asymQ{\infty} 
  &= \QvsC{+}{\infty}+ \QvsC{-}{\infty}\label{eq:asymQ}
  \\
  &= \int_{\mathrlap{-\infty}}^{\mathrlap{\infty}}\, dk\, \ab[\ab|R_{+}(k)|^{2} - \ab|R_{-}(k)|^{2}] |f(k)|^{2}.
\end{align}
We note that in the limit $\sigmakwp / k_0 \to 0$ as the wavepacket becomes infinitely peaked at $k = k_{0}$, the observable $\QvsC{\pm}{\infty}$ measures the reflection coefficients $\ab|R_{\pm}(\kwp)|^{2}$,
\begin{align}
	\QvsC{\pm}{\infty} \to \pm \ab| R_{\pm}(\kwp) | ^{2} \quad\text{as}\quad { \frac{\sigmakwp}{k_{0}} \to 0}.
\end{align}

\minisec{Phase asymmetry}
While the above observable captures any $\Csym$-asymmetry in the magnitude $\ab|R_{\pm}(k)|$, asymmetry will also lie in the phase $\phi_{\pm}(k) =  \arg R_{\pm}(k)$.
As we will see in our lattice simulations, this becomes especially important when the fractional $\Csym$-asymmetry in magnitude $\frac{\ab|R_{+}|^{2} - \ab|R_{-}|^{2}}{\ab|R_{+}|^{2} + \ab|R_{-}|^{2}} $ is too small to be detected within numerical uncertainties. In that case, there may still be a significant asymmetry in the phase.

\begin{figure}[ht]
  \centering
  \includegraphics[width=0.7\linewidth]{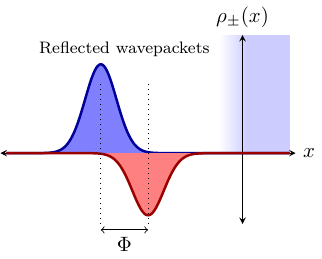}
  \caption{The phase asymmetry observable $\asymPf$ [\cref{eq:asymP}] captures the relative spatial displacement between the particle and antiparticle wavepackets, while the magnitude asymmetry $\asymQf$ [\cref{eq:asymQ}] captures the asymmetry between the total amount of charge reflected. In this picture of the local charge densities $\rho_{\pm}(x)$ for the particle (blue) and antiparticle (red) wavepackets, $\asymPf$ is the distance between the peaks of the Gaussian wavepackets, and $\asymQf$ is the sum of (signed) areas enclosed by the wavepackets.%
    \label{fig:asymmetry}}
\end{figure}

An observable that is sensitive to the phase of $R_{\pm}(k)$ is the spatial displacement of the reflected wavepackets, shown schematically in \cref{fig:asymmetry}.
We provide a more careful explanation in \cref{sec:observables}, but the essential idea of this observable can be simply understood as follows.
For a relativistic Gaussian wavepacket sharply peaked around $k = \kwp$, a given $k$ wave transforms after reflection as,
\begin{align}
  e^{ik(x - t)} 
  &\to  
  |R_{\alpha}(k)|\, e^{-ik[x+t] -i\phi_{\alpha}(k)} \\
  &\approx 
  |R_{\alpha}(k)|\, e^{-ik[x+t + \phi'_{\alpha}(\kwp)]}
\end{align}
where we expand the phase $\phi_\alpha(k)$ about $k = \kwp$ as
$\phi_\alpha(k) = \phi_\alpha(\kwp) + (k - \kwp) \phi'_\alpha(\kwp) + \Order(k^2)$ for $\alpha = \pm$,
and drop the $k$ independent phase $\phi_\alpha(\kwp)$.
We see that the reflected wave at a given time $t$ gets displaced by an amount $\phi'_{\alpha}(\kwp)$ relative to the case with a trivial phase-shift $\phi_\alpha(k) = 0$.
Therefore, we may use the difference in the displacement of the sharply peaked Gaussian wavepackets for the particle and antiparticle after reflection as a measure of the phase asymmetry:
\begin{align}
  \asymPf = \phi'_{+}(\kwp) - \phi'_{-}(\kwp).
  \label{eq:asymP}
\end{align}
Assuming that the reflected wavepacket is itself well approximated by a Gaussian, we can relate $\asymPf$ to the correlation between the charge densities $\rho_{\pm}(x,t)$ [defined in \cref{eq:rho-pm}].
At large times $t \to \infty,$ we find
\begin{align}
	\lim_{t \to \infty} \int_{\mathrlap{-\infty}}^{0} dx\, \rho_{+}(x, t) \rho_{-}(x, t) = -A e^{- B \asymPf^{2}}
\end{align}
for certain constants $A,B$  independent of $\asymPf$, where the integration limits ensure that only the reflected wavepackets contribute.
This relation motivates us to \emph{define} a new (time-dependent) phase-asymmetry observable,
\begin{align}
\asymP{t}  = \sqrt{-\frac{1}{B(t)} \log \ab(-\frac{1}{A(t)}
\int_{\mathrlap{-\infty}}^{\mathrlap{0}} dx\, \rho_{+}(x,t) \rho_{-}(x,t) )},
  \label{eq:asymP-cont}
\end{align}
where $A(t), B(t)$ are now time-dependent functions that do not depend on cross-correlations of $\rho_{\pm}$. 
In the limit $t \to \infty$, we get $\asymP{t} \to \asymPf$ of \cref{eq:asymP}.
See \cref{sec:observables} for a more detailed discussion of this observable, including the definitions of $A(t), B(t)$. 

\subsection{Mass profiles}

To complete the description of our simulations, we need to specify the mass profiles used for the bubble wall $m(x)$.
For \ac{EWBG} calculations, a commonly used mass profile is of the hyperbolic form \cite{Ayala:1993mk}, given by:
\begin{align}
|m(x)| &= \frac{m_0}{2} \ab[ 1+\tanh(x/\Lw) ], \nonumber\\
\theta(x)   &= \frac{\theta_0}{2} \ab[ 1+\tanh(x/\Lw) ],
\label{eq:massprofile}
\end{align}
where $\Lw$ is the width of the bubble wall.
The parameter $\theta_{0}$ controls the degree of $\Csym$ violation.
In this work, we consider two cases.

First, we consider a real mass profile, $\theta_0 = 0$ with no $\Csym$-violation.
The Dirac equation in this case is, in fact, analytically solvable \cite{Funakubo:1994en} and thus serves as a benchmark for our lattice simulations.
The reflection coefficient $\ab| R_{+}(k) |^{2} = \ab| R_{-}(k) |^{2} $ is \cite{Funakubo:1994en}
\begin{equation}
  \ab| R_{\pm}(k) |^{2} = \begin{cases}
    \frac{ \sin\ab[\frac{\pi}{2}(\beta' -\beta+\xi)] \sin\ab[\frac{\pi}{2}(\beta'-\beta-\xi)] }{ \sin\ab[\frac{\pi}{2}(\beta'+\beta+\xi)] \sin\ab[ \frac{\pi}{2}(\beta'+\beta-\xi)]}, &  k>m_0\\
    1,   &   k\leq m_0
    \end{cases},
    \label{eq:Rk-tanh-analytic}
\end{equation}
where $\beta'=i\sqrt{\epsilon^2-\xi^2}$, $\beta=i\epsilon$, with the dimensionless quantities $\epsilon\equiv k \Lw$, $\xi\equiv m_0 \Lw$.
As there is no $\Csym$ symmetry breaking, the asymmetry observables should vanish: $\asymQf = 0$ and $\asymPf = 0$.

We then consider the nontrivial case of a complex mass with $\theta_0\neq 0$, where no analytical solution is available.
A complex mass term produces an asymmetry in the reflected particle and antiparticle wavepackets.
The goal of this work is to show the above mentioned asymmetry observables can be computed using real-time simulations.


%% file: sec_lattice.tex
\section{Lattice Simulations}
\label{sec:lattice}

We now translate the continuum setup described in the previous section to the lattice to enable a non-perturbative study of $\Csym$-asymmetry generation during fermion-bubble collisions.
We then describe the results of the simulations performed using \ac{MPS} \cite{itensor}.

\subsection{The lattice setup}
\label{sec:lattice-setup}

\newcommand\HLat{\hat{H}}
\minisec{Hamiltonian}
Staggered fermions \cite{Kogut:1974ag} provide a convenient doubler-free lattice formulation for a Dirac fermion in $1+1$ dimensions by mapping the two spinor components of the Dirac field $\psi(x)$ in \cref{eq:hcon} to the odd and even sites.
On a $N$-site ($N$ even) chain with \ac{OBC}, the staggered fermion Hamiltonian is
\begin{align}
  \HLat &= \sum_{n=1}^{N-1}  i \ab[ \frac{1}{2} + (-1)^n |m_n| \sin\theta_n ](\chi_{n+1}^\dagger\chi_n^{\mathstrut}-\chi_n^\dagger\chi_{n+1}^{\mathstrut})
        \nonumber\\
  &\quad -  \sum_{n=1}^{N}(-1)^n |m_n| \cos\theta_n \, \chi_n^\dagger\chi_n^{\mathstrut},
    \label{eq:hpos}
\end{align}
where $\chi_{n}$  are 1-component complex fermion fields satisfying $\{ \chi_i^{\dagger}, \chi_{j} \} = \delta_{ij}$.

We use $m_n\equiv  a m(an - a\Nc)$ and $\theta_{n} \equiv \theta(an- a\Nc)$ as discretizations of the mass profile given in \cref{eq:massprofile}, where $\Nc = (N+1) / 2 $ is the center of the bubble wall.
The symmetric phase can thus be identified as the half space of sites $n$ with $1 \leq n \leq N / 2$, and the broken phase as the other half with $N / 2 +1 \leq n \leq N$. 

\minisec{Initial state and time evolution}
We use \ac{DMRG} to compute the ground state $|\vacLat\>$ \cite{white_density_1992,schollwock_densitymatrix_2011}.
Despite the vanishing mass gap in the symmetric phase, we find that \ac{DMRG} converges well for the system sizes considered in this work.

Let ${\psiLat}^{\dagger}_{k \alpha}$ be the creation operators on the lattice for particles ($\alpha =+$) and antiparticles ($\alpha=-$) with momentum $k$ for the staggered-fermion Hamiltonian in \cref{eq:hpos}.
(See \cref{sec:create-wp} for precise definitions on the lattice.)
Strictly speaking, lattice momenta $k$ are not well-defined in \ac{OBC}.  However, we define the single-particle momentum eigenstates in \ac{PBC} and assume that the volumes are large enough such that the errors are negligible.  We study the finite-size effects in more detail in \cref{sec:systematics}.

In analogy with the continuum formalism discussed in the previous section, we define the wavepacket initial states
\begin{align}
  \ket| \Psi_\pm(0)> = \ket| f_\pm > = \mathcal{N} \sum_{\kLat} f(k) {\psiLat}^{\dagger}_{k \pm} \ket| \vacLat >
\end{align}
where $\mathcal{N}$ is the normalization constant such that $\braket< f_\pm| f_\pm > = 1$,
and $\kLat$ runs over lattice momenta in \ac{PBC}, $\kLat = - \pi/2 + 2 \pi j/N $ for $j=1,\dotsc,N/2$.
We place the center of the wavepacket at $\xwpLat = N / 4$ to have minimal overlap with the bubble wall and the lattice boundary.
The exact construction of the wavepacket is specified in \cref{sec:create-wp}.

The state $\ket| \Psi_\pm(t)>$ is then obtained by time evolution with the Hamiltonian in \cref{eq:hpos}.
The real-time evolution can be computed using a simple second-order Trotter scheme as described in \cref{sec:create-wp}.

\minisec{Choice of parameters}
In the continuum theory, the mass $m_{0}$ of the fermion in the broken phase is the natural physical scale.
Therefore, in the following, we report all quantities in physical units of $m_{0}$.

We fix the lattice volume to be $\Lphys=28$ and the wall width to be $\Lw = 0.6$ for all simulations in this work.
We perform time evolution for a total time $T = \Lphys$. For this,
we use a Suzuki-Trotter scheme with the step size fixed at $\tau = 0.05 $,
which is much smaller than the inverse of the momentum resolution $\Delta k = 2\pi/L \approx 0.22$.
We find that the trotter errors are negligible compared to finite-size and lattice artifacts, so we do not perform $ \tau \to 0$ extrapolations.

One of the most challenging aspects of these simulations is that we need the wavepacket to be localized in both position and momentum space.
This results in the presence of multiple scales that need to fit in the box.
The initial Gaussian wavepacket is characterized by the physical value of the central momentum $k_{0}$, the momentum width $\sigmakwp$, the central position $x_{0}$, and the spatial width $\sigma_x=1/(2\sigmakwp)$.
With the bubble wall located at $x = 0$,
localization of the initial wavepacket in position space implies
\begin{align}
  0 \ll \ab| \xwp \pm \sigmaxwp|  \ll \frac{\Lphys}{2}.
\label{eq:scale-x}
\end{align}
On the other hand, we need the wavepacket to also be sufficiently localized in momentum space to avoid lattice artifacts from large momenta, and to avoid the presence of negative-momentum (left-moving) modes:
\begin{align}
 0 \ll \ab|\kwp \pm \sigmakwp| \ll \frac{\pi}{2a}.
\label{eq:scale-k}
\end{align}
We take $\xwp = -\Lphys/4$ so that the initial wavepacket has a minimal overlap with the left boundary at $x = -L/2$ and the bubble wall at $x = 0$.
Noting that $\sigmaxwp \sigmakwp =1/2$, the above inequalities can be satisfied if we choose the wavepacket momentum space parameters $\kwp, \sigmakwp$ such that
\begin{align}
  \frac{\pi}{2 a } \gg \kwp  \gg \sigmakwp  \gg \frac{2}{\Lphys}.
  \label{eq:scale-sep}
\end{align}
Getting a clean separation of all these scales is a numerical challenge for any lattice simulation.
In particular, for a fixed $\sigmakwp$, we expect there to be increased systematic errors as we make $\kwp$ too small ($\kwp \lesssim \sigmakwp$), as well as when $\kwp$ becomes large ($\kwp \gtrsim {\pi}/{2a}$).

We vary the lattice spacing $a$ in the range $\{\frac{1}{8}, \frac{1}{6}, \frac{1}{5}, \frac{1}{4}\}$ for a continuum limit extrapolation. This corresponds to a lattice size $N = L / a$ varying in the range $N = 112, \dotsc, 224$.
The largest lattice spacing we have is thus $a = 0.25 $.
Therefore, \cref{eq:scale-sep} requires that $6.28 \gg \kwp \gg \sigmakwp \gg  0.07$.
Ideally, we would like $\sigmakwp$ to be as small as possible to allow maximum variation in $\kwp$.
However, we are also limited by the resolution in momentum space $ \Delta k = 2\pi / L \approx 0.22$.
So we take $\sigmakwp = 0.5 \approx 2 \Delta k$ and keep it fixed for all simulations.\footnote{For a study of finite-volume effects, see \cref{sec:systematics}.}

 \minisec{Observable (i) Magnitude Asymmetry}
 Defining the local charge density of the states $\ket| \Psi_\pm(t)>$,
 \newcommand\rhoLat[3]{\rho^{#1}_{#2}(#3)}
 \begin{align}
   \rho_{n, \pm}(t)
   = \braket< \Psi_\pm(t) | \jvLat^{0}_{n} | \Psi_\pm(t)> -  \braket< \vacLat | \jvLat^{0}_{n} | \vacLat > ,
   \label{eq:rho-lat}
 \end{align}
 where we subtract the vacuum contribution to reduce lattice artifacts,
 we define the total charge in the symmetric phase on the lattice as the lattice analog of \cref{eq:qvst},
\begin{align}
  \QvsL{t}{\pm}
  &= \sum_{n=1}^{N}
    w_{n} \rho_{n, \pm}(t)
    \label{eq:QvsL}
\end{align}
where $\jvLat^0_{n} = \chi_n^{\dagger}\chi_n^{\xdagger}$ is the $U(1)$ charge density on the lattice at site $n$, and $w_{n}$ is a suitably chosen ``smeared'' step function which vanishes deep in the broken phase $w_{n} = 0$ for $n\gg \Nc$, and is unity in the symmetric phase: $w_n = 1$ for $n \ll \Nc$.
A smooth $w_{n}$ reduces contributions close to the wall, thereby reducing lattice artifacts.
For the measurement of the magnitude asymmetry, we choose $w_n$ to be the Sigmoid function:
\begin{align}
    w_n = \ab[\exp\ab(\frac{n-(N_c-n_w)}{0.1\,n_w})+1]^{-1},
    \label{eq:wfunc_Qs}
\end{align}
where $n_w=\Lw/a$ is the wall width in lattice unit.

\begin{figure*}
  \includegraphics[width=\linewidth]{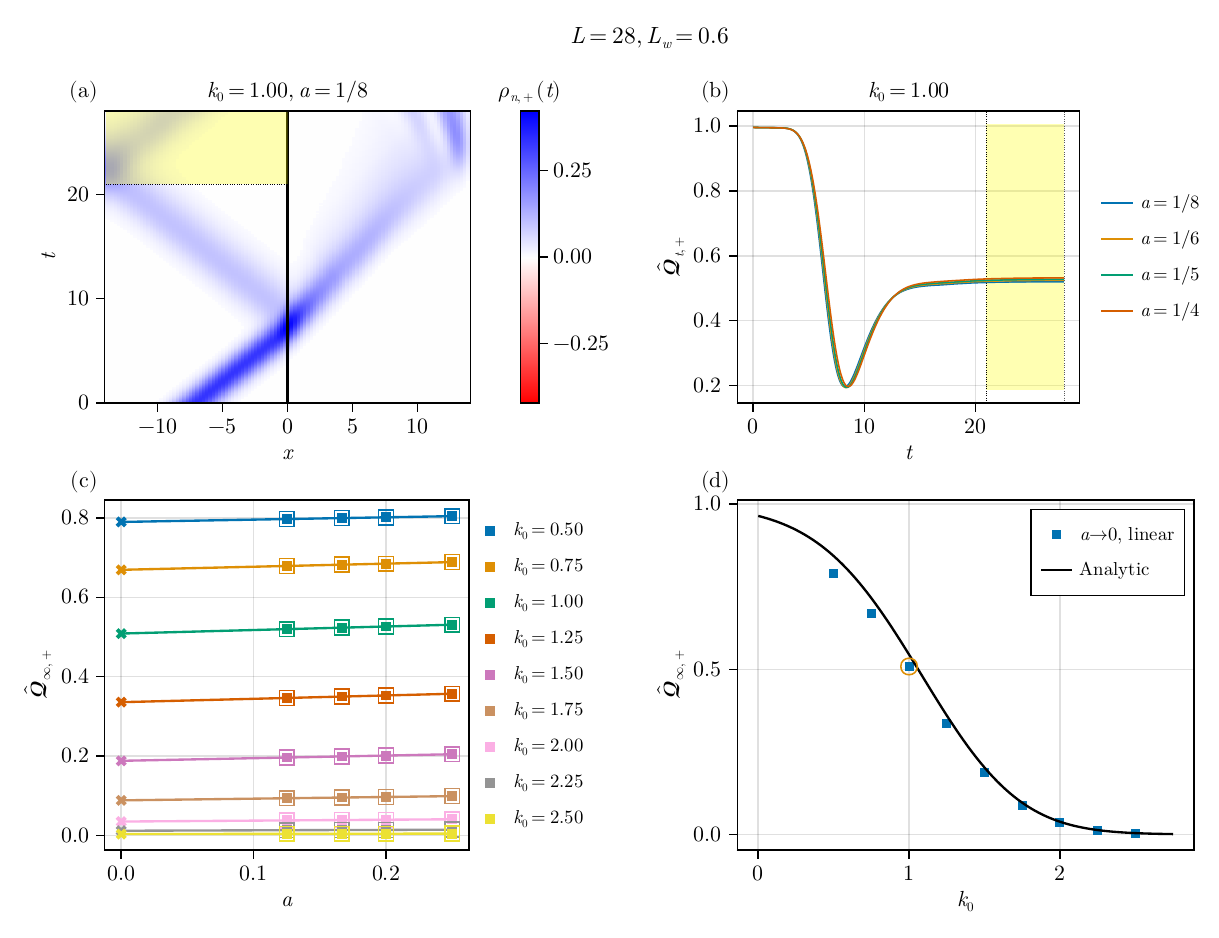}
  \caption{
    Simulation results for a fermion wavepacket with $\sigmakwp = 0.5$ scattering off a real mass profile with wall width $\Lw=0.6$ in a lattice box of size $\Lphys = 28$.
    (a) Evolution of the particle number density over the lattice for $\kwp=1.0$ and $a=1/8$.
    (b) Particle number in the symmetric phase, $\QvsL{t}{+}$, as function of time at different lattice spacings.
    The vertical dotted line shows the time $t = 3 T / 4$.
    The yellow highlighted band in (a) and (b) shows the time window $\tWindowQ$ used for the asymptotic value $\QvsLf+$.
    (c) Continuum limit extrapolations for $\QvsLf{+}$  using linear fit for various values $\kwp$.
    The measurement errors are too small to be visible on this scale.
    (d) Continuum limit of $\QvsLf{+}$ as a function of $\kwp$, compared with the analytic prediction (black line) $R_{f}(\kwp)$ [\cref{eq:R-analytic-wavepacket}]. The result for $\kwp = 1$, corresponding to panels (a) and (b), is highlighted by a yellow circle.
  }\label{fig:real-mass}
\end{figure*}

The magnitude asymmetry observable on the lattice is then
\begin{align}
  \asymQL{t} = \QvsL{t}{+} + \QvsL{t}{-}
  \label{eq:asymQL}
\end{align}
In the infinite-volume limit, $\asymQL{t}$ saturates to the total reflected charge $\asymQLf$ as $t \to \infty$.
However, this is not the case in a finite box, and therefore we need to choose an appropriate time window to measure this observable with minimal finite-size errors.
The finite size effects will be minimized when the reflected wavepacket has minimal overlap with the bubble wall.
Since the wavepacket in the symmetric region travels with a velocity $v = 1$, and the wavepacket is initially located at the center of the symmetric region, we choose to measure the observable in the window $t \in \tWindowQ \equiv (3 L /4, L)$.
This is when the reflected wavepacket is expected to hit the left boundary and return to its starting position.
We use the mean value of $\asymQL{t}$ in the window $\tWindowQ$ as the measurement, and use the standard deviation of $\asymQL{t}$ over the time window $\tWindowQ$ as an estimate of the systematic errors due to finite volume.
[See \fig{real-mass}{a} for an illustration of the time window $\tWindowQ$, which is highlighted as a yellow band.]

\minisec{Observable (ii) Phase Asymmetry}
We use a naive discretization of the phase asymmetry observable $\asymP{t}$ in \cref{eq:asymP-cont} given by
\begin{align}
  \asymPL{t} =	\sqrt{-\frac{1}{B(t)} \log\ab[-\frac{1}{A(t)} \sum_{n=1}^{N}a^{-1} w'_{n} \rho_{n,+}(t) \rho_{n,-}(t)]}.
  \label{eq:asymPL-t}
\end{align}
where $w_{n}'$ is again a suitably chosen smearing function that vanishes in the broken phase.
We choose $w_{n}'$ to be a double-sided sigmoid function to suppress the boundary effects from both the bubble wall and the lattice boundary (see discussion in the next paragraph):
\begin{align}
    w_n'=w_n\ab[\exp\ab(-\frac{n-n_w}{0.1\,n_w})+1]^{-1},
\end{align}
with $w_n$ given by \cref{eq:wfunc_Qs}.
See the discussion following \cref{eq:asymP-AB} for the expressions of $A(t)$ and $B(t)$ on the lattice. 

\begin{figure*}[htbp]
    \includegraphics[width=\linewidth]{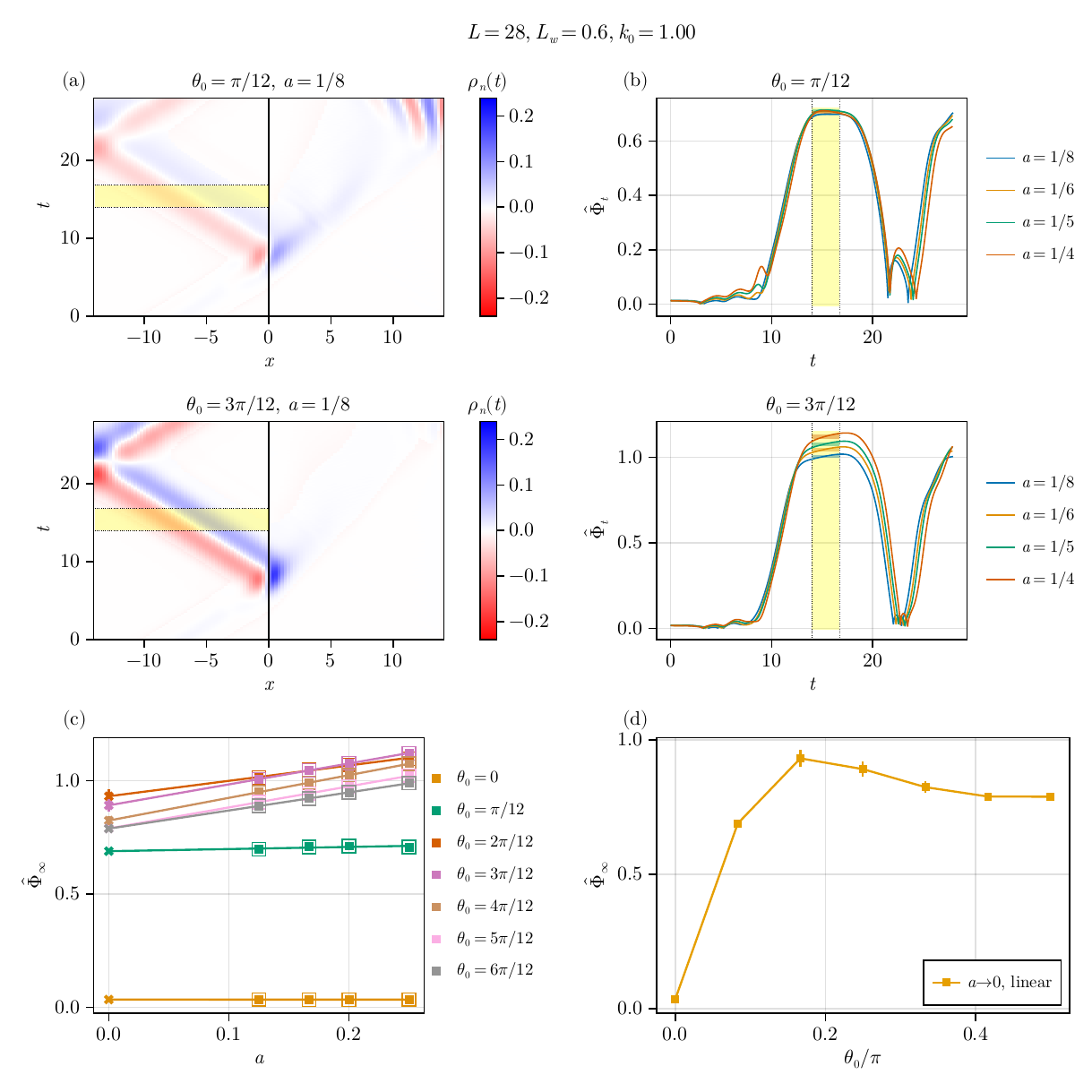}
    \caption{
      \emph{Phase asymmetry} as a measure of $\Csym$-asymmetry generated
      during fermion and antifermion scattering with a complex mass profile for fixed wavepacket with $\kwp = 2\sigmakwp = 1.0$ and physical volume of $\Lphys = 28$.
      For two parameters $\theta_0 = \pi / 12, 3 \pi /12$,
      (a) shows the net charge density in spacetime, and (b) the evolution of the phase asymmetry in the symmetric phase $\asymPL{t}$ at different lattice spacings $a$. The yellow band marks the time-window $\tWindowP$ used to measure the asymptotic value $\asymPLf$ and estimate the systematic uncertainties.
      (c)
      Continuum limit extrapolations for $\asymPLf$ using linear fit for various values of $\theta_0$.
      The errors from the measurements and the $1\sigma$ errors from the fitting are shown for the finite $a$ points and the extrapolated $a\to 0$ point, respectively.
      (d) Finally, $\asymPLf$ as function of $\theta_0$.
      The error bars are the $1\sigma$ uncertainties computed in $a \to 0$ fits of panel (c).
}
    \label{fig:phase-asym}
\end{figure*}

The choice of a suitable time window $\tWindowP$ to measure the phase asymmetry is more subtle than for the magnitude asymmetry.
This is because while the magnitude asymmetry only depends on the total charge reflected back, the phase asymmetry assumes that the wavepackets are Gaussian.
When there is significant overlap with either the bubble wall or lattice boundary, the shape of the wavepacket cannot be approximated as a Gaussian anymore.
However, if the wavepacket is sufficiently away from both boundaries, $\asymPL{t}$ becomes independent of $t$.
Thus, we choose the window $\tWindowP$ by looking for a plateau in $\asymPL{t}$.
We find that an optimal time window is approximately $\tWindowP \equiv (0.5 L, 0.6L)$
for the choice of parameters considered in this work [see \fig{phase-asym}{b} for an illustration of $\tWindowP$, highlighted as a yellow band].
This is consistent with our expectation that the reflected wavepacket is in the center of the symmetric region, and therefore is maximally away from the bubble wall and lattice boundary, at approximately $t = L / 2$.
We use the mean value in this window as the measured asymptotic value $\asymPLf$ and the standard deviation as a measure of systematic errors from the finite volume.

\subsection{Simulation Results: Real Mass ($\theta_{0} = 0$)}
\label{sec:results-real}

We first consider a fermion that scatters off a bubble wall given by a real mass profile $\theta_0 = 0$.
When $\theta_{0} = 0$, the Hamiltonian preserves $\Csym$ and therefore the particle and antiparticle wavepackets propagate identically.
We therefore only consider the particle wavepacket.
For the real mass profile in \cref{eq:massprofile}, we have an analytic computation of the reflection coefficient [\cref{eq:Rk-tanh-analytic}].
Therefore, this computation serves as an important benchmark for the validation and estimation of systematic uncertainties when we look at $\Csym$-asymmetry generation with a complex mass in \cref{sec:results-complex}.

\Cref{fig:real-mass} shows the results of the simulation for a total time of $T = \Lphys = 28$ .
In \fig{real-mass}{a}, we illustrate the time evolution of a particle wavepacket prepared with $\kwp = 1.0, \sigmakwp = 0.5$ at the lattice spacing $a = 1/8$, where the color gradient indicates the particle number density $\rho_{n, +}(t)$.
The fermion wavepacket moves towards the bubble wall with a constant velocity $v = 1$ until it hits the bubble wall at $t \sim L /4  = 7$. 
Part of the wavepacket is reflected back into the symmetric phase, while the rest is transmitted across into the massive phase. 
Around $t \sim 3 L /4 = 21$, the reflected wavepacket hits the lattice boundary and reflects again.
During this simulation, we measure the total charge in the symmetric phase $\QvsL{t}{+}$ as a function of time $t$ to keep track of the reflected wavepacket.

In \fig{real-mass}{b}, we show the behavior of $\QvsL{t}{+}$ for the same $\kwp = 1.0$ for various lattice spacings $a$.
The initial value of $\QvsL{t}{+}=1$ is maintained until the wavepacket hits the bubble wall.
After the reflection is complete, $\QvsL{t}{+}$ approaches an asymptotic value $\QvsLf{+} < 1$.
The yellow highlighted band marks the window of time $\tWindowQ$ that we use to obtain the asymptotic value $\QvsLf+$.
The variation in this time window is a measure of the finite-volume effects.
We then take the measured value $\QvsLf{+}$ as a function of $a$ and extrapolate to $a \to 0$.

The continuum limit extrapolation for $\QvsLf{+}$ is shown in \fig{real-mass}{c} for $k$ in the range of $[0.5, 2.5]$ in steps of $0.25$.
We find that a two-parameter linear fit (shown by a solid line) works well.
The uncertainties from the variation of the observable in the time window are too small to be seen on this scale.

Finally, we show a comparison between the lattice simulations and the analytic predictions for $\QvsLf{+}$ as a function of the wavepacket momentum $\kwp$ in \fig{real-mass}{d}.
The continuum-extrapolated values of $\QvsLf{+}$ are shown by filled squares.
The continuum, infinite-volume, analytic prediction (shown by a solid line) is computed using $|R_{+}(k)|^{2}$ from \cref{eq:Rk-tanh-analytic} with the wavepacket $f(k; \kwp, \sigmakwp, \xwp)$ in \cref{eq:wp-wavefunction} as
\begin{align}
  {R}_{f}(\kwp) = \int_{0}^{\mathrlap{\infty}} dk\ | R_{+}(k) f(k;  \kwp, \sigmakwp, \xwp) |^{2},
  \label{eq:R-analytic-wavepacket}
\end{align}
where we set $\sigmakwp = 0.5$ for the incoming wavepacket, and the above expression is independent of $\xwp$.
We observe a good agreement between the lattice and the analytic predictions. However, compared to the error bars from the continuum limit fits, which are too small to be visible on the plot, the deviation between the lattice and analytic results is much larger.
We observe larger deviations with respect to the error bars for smaller $k_0$, which is expected from a consideration of scale separation.
At a fixed $\sigmakwp = 0.5$, the small momentum $\kwp \sim \sigmakwp$ invalidates the scale separation of \cref{eq:scale-sep}. This results in a shape distortion of the wavepacket away from a Gaussian, since we do not allow negative $k$ modes in the wavepacket.
Ideally, we could make $\sigmakwp$ smaller, but that would require a larger physical volume to fit a wider wavepacket in position space.
So, the disagreement at lower momentum is a systematic error from finite-volume effects.

\begin{figure*}[htbp]
    \includegraphics[width=\linewidth]{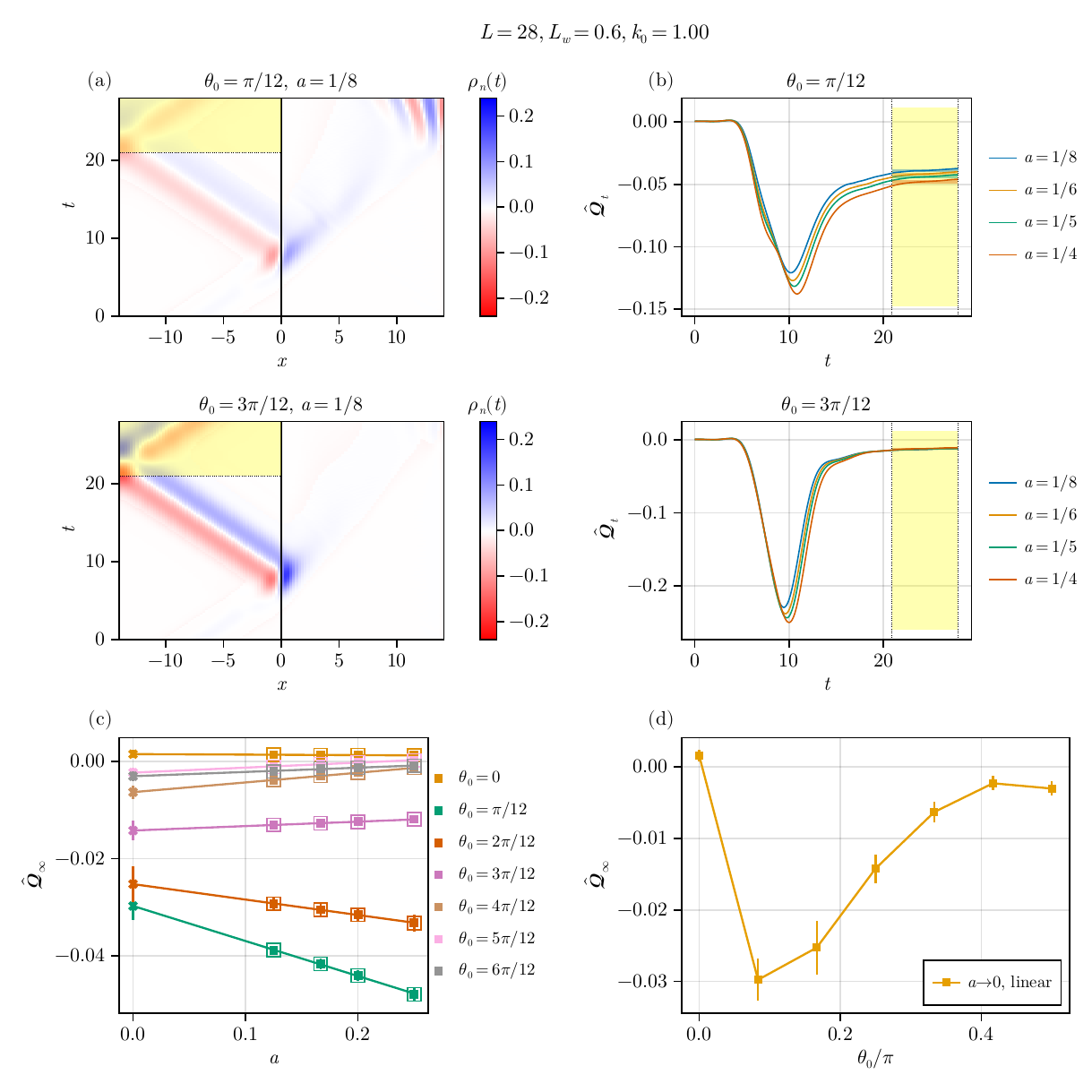}
    \caption{
      \emph{Magnitude asymmetry} as a measure of $\Csym$-asymmetry generated
      during fermion and antifermion scattering with a complex mass profile for fixed wavepacket with $\kwp = 2\sigmakwp = 1.0$ and physical volume of $\Lphys = 28$.
      For two parameters $\theta_0 = \pi / 12, 3 \pi /12$, (a) shows the net charge density in spacetime, and (b) the evolution of the net charge density in the symmetric phase $\asymQL{t}$ at different lattice spacings $a$.
      The yellow band marks the time-window $\tWindowQ$ used to measure the asymptotic value $\asymQLf$ and estimate the systematic uncertainties, as described in \cref{sec:lattice}.
      (c)
      Continuum limit extrapolations for $\asymQLf$ using linear fit for various values of $\theta_0$.
      The errors on the measurements are too small to be seen on this scale, and the $1\sigma$ errors from the fitting are shown for the extrapolated $a\to 0$ points.
      (d) Finally, $\asymQLf$ as function of $\theta_0$.
      The error bars are $1\sigma$ uncertainties computed in $a \to 0$ fits of panel (c).
    }
    \label{fig:mag-asym}
\end{figure*}

\subsection{Simulation Results: Complex Mass ($\theta_{0} \neq 0$)}
\label{sec:results-complex}

Having validated our methods with an estimate of the expected uncertainties, we now proceed to the nontrivial case of a complex mass profile where a genuine $\Csym$-asymmetry generation may be observed.

For studying the $\Csym$-asymmetry, we independently simulate the particle and antiparticle scattering processes and combine the results.
We can equivalently formulate our discussion in terms of the time evolution of a completely mixed state $\rho_{0} = \sum_{\alpha = \pm} \ket| f_{\alpha} > \bra< f_{\alpha} |$
of the $\Csym$-conjugate pair of wavepacket states $\ket| f_\pm >$.
As this initial mixed state is symmetric under $\Csym$, observables that are not invariant under $\Csym$ can be used to measure the generation of $\Csym$-asymmetry in collision.
However, for efficiency of the tensor network algorithms, we perform separate simulations for the $\Csym$-conjugate pure states $\ket| f_{\pm} >$.

We would like to study the asymmetry generation for the mass profile \cref{eq:massprofile} as a function of $\theta_{0}$.
For these simulations, we vary $\theta_{0}$ in the range $[0, {\pi} / {2}]$ in steps of ${\pi} / {12}$.
The wavepacket parameters are fixed to $\kwp = 2\sigmakwp = 1.0$.
All other parameters are as before.
We now describe the results for both the magnitude and phase asymmetry observables.

\newcommand\rhoLatNet[2]{\rho_{#1}(#2)}

\minisec{Magnitude asymmetry}
The results for the magnitude asymmetry are shown in \cref{fig:mag-asym}.
In panels (a) and (b), we focus on two values of $\theta_{0}$ for illustration $\theta_0 = \pi /12, 3 \pi / 12$.
In \fig{mag-asym}{a}, we show the net local charge density,
\begin{align}
  \rhoLatNet{n}{t} \equiv \rhoLat{}{n, +}{t} + \rhoLat{}{n, -}{t},
  \label{eq:rho-lat-net}
\end{align}
for lattice spacing $a = 1/8$. 
For both $\theta_0$s, the $\Csym$-asymmetry can be clearly seen from the nonvanishing $\rhoLatNet{n}{t}$.
In \fig{mag-asym}{b}, we plot the time evolution of the magnitude-asymmetry observable $\asymQL{t}$ [\cref{eq:asymQL}], measuring the net charge in the symmetric region, for various lattice spacings $a$.
The total charge in the symmetric region $\asymQL{t}$ is zero at $t = 0$, since the initial wavepacket states are $\Csym$ conjugates of each other.\footnote{The initial wavepacket states are actually not exact $\Csym$-conjugates of each other in the presence of a complex mass due to a finite wavepacket width, which introduces a small systematic error from the overlap of the initial wavepacket with the bubble wall. We examine this in \cref{sec:errors-Csym-initial}.}
After an initial period of no change, the wavepacket scatters from the bubble wall around $t \sim 7$, where we see a large dip in $\asymQL{t}$.
Around $t \sim 21$, the reflected wavepacket has minimal overlap with the bubble wall, and therefore $\asymQL{t}$ begins to saturate.
The yellow band therefore marks the time window $\tWindowQ$ which we used to define the asymptotic value $\asymQLf$, as discussed previously in \cref{sec:lattice-setup}.

\fig{mag-asym}{c} shows the continuum limit $a \to 0$ extrapolation of $\asymQLf$ for the entire range of $\theta_{0} $.
We find that a linear fit works well. The errors in the measurements are too small to be visible, but the error of the linear fit $a\to 0$ can be seen on the $a=0$ axis. We observe that for $\theta_0 = 0$, the central value of $\asymQLf$, which is expected to be zero, is comparable to the systematic errors of the initial state preparation as we studied in \cref{sec:errors-Csym-initial}.
Interestingly, we find that the lattice artifacts, and therefore the total systematic errors as shown in \fig{mag-asym}{d}, are larger for intermediate $\theta_0$ values.

Finally, we show the continuum-extrapolated magnitude asymmetry $\asymQLf$ as a function of $\theta_{0}$ in \fig{mag-asym}{d}.
While these results establish $\asymQLf$ as an easily measurable observable quantifying the $\Csym$-asymmetry generated during the fermion-bubble collision,
they also demonstrate a clear limitation of this observable. 
At large $\theta_{0}$, $\asymQLf$ is small simply because of the strong reflection but not because the asymmetry itself is small.
Because both the particle and the antiparticle are completely reflected, the net reflected charge is almost zero.
But the local charge density plot in panel \fig{mag-asym}{a} clearly shows a large asymmetry in the relative displacement of the particle and antiparticle wavepackets even for the larger $\theta_{0} = 3 \pi /12$.
This demonstrates that when the reflection coefficient approaches unity, $|R_{\alpha}(\kwp)|^{2} \approx 1$, magnitude asymmetry no longer serves as an effective measure of $\Csym$-asymmetry.
Consequently, we turn our attention to the phase asymmetry in $R_{\alpha}(\kwp)$ in what follows.

\minisec{Phase asymmetry}
We now look at the phase asymmetry observable $\asymPL{t}$ defined in \cref{eq:asymPL-t}, with the results shown in \cref{fig:phase-asym}. \fig{phase-asym}{a} shows the same net local charge density $\rhoLatNet{n}{t}$ as previously shown in \cref{fig:mag-asym}. However, the time window $\tWindowP$ over which $\asymPLf$ is measured is different, shown as a yellow band in \fig{phase-asym}{a}.
The reason for this can be seen from \fig{phase-asym}{b}, where we plot $\asymPL{t}$ as a function of $t$ for various lattice spacings $a$.
Our definition of $\asymPLf$ assumes a Gaussian wavepacket, which is valid only when the reflected wavepacket is sufficiently far from any boundaries.
The presence of a plateau in \fig{phase-asym}{b} shows that the yellow band is a reasonable choice for $\tWindowP$.
In the limit of $\Lphys/\sigma_x \to \infty$, \fig{phase-asym}{b} would be completely flat in this time window.
Therefore, any variation in this window is a finite-volume effect.

In \fig{phase-asym}{c}, we show the continuum limit extrapolation $a \to 0$ for $\asymPLf$ for various $\theta_{0}$.
Again, we find that a linear extrapolation works well.
The dependence of phase asymmetry $\asymPLf$ (in the $a \to 0$ limit) on the parameter $\theta_{0}$ is shown in \fig{phase-asym}{d}.
At $\theta_{0} = 0$, we observe a small non-vanishing $\asymPLf$ which is again comparable to the systematic errors in $\asymPLf$ from initial state preparation, as discussed in \cref{sec:errors-Csym-initial}.
For large $\theta_{0}$, the phase-asymmetry observable reveals a significant $\Csym$-asymmetry, which could not be detected by the magnitude-asymmetry observable.

The behavior of the instantaneous charge asymmetry in \fig{phase-asym}{a}, shown by the non-vanishing $\rho_n (t)$, in real time near the wall, merits particular attention. In \ac{EWBG}, $SU(2)$ sphalerons active in the symmetric phase can convert the analogous $\CP$ asymmetry in 3+1 dimensions to baryon asymmetry.
As the bubble expands, the baryon asymmetry generated near the wall penetrates into the broken phase, where it remains preserved when electroweak sphalerons become inactive for a sufficiently strong phase transition.
This suggests that the instantaneous $\CP$ asymmetry generated at the collision point near the wall may play a crucial role in explaining the observed baryon asymmetry of the universe, necessitating careful consideration in future nonperturbative studies.


%% file: sec_conclusion.tex
\section{Conclusions}
\label{sec:conclusions}

Motivated by the goal of a fully nonperturbative quantum simulation of baryon-number asymmetry generation during \ac{EWBG}, we initiated the study of $\Csym$-asymmetry generation during fermion-bubble collisions in 1+1 dimensions.
Although quantum technologies evolve and a full-scale simulation of \ac{EWBG} remains beyond immediate reach, tensor network methods provide an extremely useful way to prototype future quantum simulations in lower dimensions.

In this first study, we explored a toy model consisting of free fermions moving in the rest frame of a bubble wall modeled as a complex mass profile.
The particle and antiparticle wavepackets reflect off the bubble wall differently, which is the source of the $\Csym$ asymmetry generated on the symmetric side.
Emphasizing the need to carefully design real-time observables that can capture the $\Csym$-asymmetry, we proposed two observables constructed from local charge densities of the final states.
The first is the magnitude asymmetry, which measures the total net charge in the symmetric phase after scattering, and the second is the phase asymmetry, which measures the relative spatial displacement between particle and antiparticle wavepackets.
In regimes of strong reflection, which occur at large $\Csym$-violating phase $\theta_0$, the simplest observable -- the magnitude asymmetry -- is too small, comparable to systematic errors, thereby creating a signal-to-noise ratio problem.
We observe a strong asymmetry in the relative displacement of the wavepackets in real time, which can be successfully captured by the phase-asymmetry observable.
This observable does not suffer from the signal-to-noise problem afflicting the magnitude asymmetry, although it does require a careful choice of measurement time window due to strong boundary effects.

Beyond applications to early-universe dynamics, our methods for computing phase asymmetry could be broadly relevant for nonperturbative computations of scattering amplitudes in \acp{QFT}, as an alternative to the Lüscher method \cite{luscher_two-particle_1991} used in lattice \ac{QCD}.
Exploring such applications in the context of two-particle scattering on quantum computers remains an important direction.
Our phase-asymmetry observable is closely related to the time-delay measurement of Ref.~\cite{gustafson_realtime_2021}, though their approach differs fundamentally from ours.

We have shown that lattice methods can compute a \emph{quantitative} measure of $\Csym$-asymmetry generation in a 1+1-dimensional toy model of baryogenesis, demonstrating how real-time simulations can quantitatively probe nonequilibrium phenomena.
We focus on a phenomenologically relevant measure of $\Csym$-asymmetry often computed using traditional analytic methods without systematically improvable uncertainties. Real-time lattice simulations instead supplement them by computing such observables in a controlled manner.
We emphasize that although we compute the asymmetry away from the bubble walls, where reflection coefficients are well defined, the methods are also applicable for computing the asymmetry closer to the bubble wall, which could in fact be the relevant quantity for precise estimates of baryon asymmetry in the broken phase.
There is no in-principle obstruction to also including thermal effects within the simulation itself, which would complement the analytic approaches which use diffusion equations.
In the context of early-universe dynamics, Refs.~\cite{milsted_collisions_2022,jha_realtime_2024} studied the scattering of dynamical bubble walls (without fermions) using tensor networks.
Combining their techniques with ours provides a natural pathway to study asymmetry generation in the presence of dynamical bubble walls.
Therefore, this work serves as a starting point for investigating ingredients such as sphalerons converting chiral asymmetry to baryon-number asymmetry, dynamical bubble walls, thermal effects, and particle transport, preparing us for a fully nonperturbative \emph{ab-initio} simulation of baryogenesis.

\section*{Acknowledgments}
We would like to thank
Vincenzo Cirigliano,
Yang Bai,
Tim Hobbs and
Henry Lamm
for many useful conversations.
TO is supported by the Visiting Scholars Program of URA at Fermilab and the DOE Office of Science Distinguished Scientist Fellows Award 2022.
The work of MC (partially) and HS was supported by the Department of Energy through the Fermilab QuantiSED program in the area of ``Intersections of QIS and Theoretical Particle Physics.''
This manuscript has been authored by the Fermi Forward Discovery Group, LLC, under Contract No. 89243024CSC000002 with the U.S. Department of Energy, Office of Science, Office of High Energy Physics.
The research of MC at Perimeter Institute is supported in part by the Government of Canada through the Department of Innovation, Science and Economic Development, and by the Province of Ontario through the Ministry of Colleges and Universities.
YYL is supported by IHEP under Grant No. E55153U1.


%% file: app_symmetries.tex
\section{Symmetries}
\label[appendix]{sec:cpt}

In this section, we discuss the symmetries of the staggered fermions formulation and their dynamics.

\subsection{Continuum}

\minisec{Parity}
The massless Dirac fermion in even spacetime dimensions has a parity symmetry $\Psym$. We define this to be a symmetry which flips the sign of all the spatial coordinates.
Under $\Psym$, we have
\begin{align}
	\Psym:
  \begin{cases}
    \psi(t,x) \to \kappa \gamma_0 \psi(t, -x) \\
    \bar\psi(t,x) \to
    \bar\psi(t, -x) (\kappa \gamma_0)^{\dagger}
  \end{cases}
\end{align}
where $\kappa = 1 $ or $\kappa = i$ such that $(\kappa \gamma_0)^{2} = 1$, so that this is valid of any choice of signature, either Minkowski or Euclidean.
One can check that this is always a symmetry of the fermion kinetic term, but not necessarily for a mass term.
A general fermion bilinear transforms under parity as
\begin{align}
	\overline{\psi} \Gamma \psi \xrightarrow{\ \Psym\ } \overline{\psi} \gamma_0^{-1} \Gamma \gamma_{0} \psi.
\end{align}

\minisec{Charge-conjugation}
There are actually two definitions of charge conjugation since
any charge conjugation symmetry can be transformed into another by a discrete chiral transformation.
To define the charge-conjugation symmetry $\Csym$, we first define unitary matrices $\Cmat_{\epsilon}$ $(\epsilon = \pm)$,
such that
\begin{align}
	\Cmat_{\epsilon} \gamma_{\mu} \Cmat^{-1}_{\epsilon} &= \ \epsilon (\gamma_{\mu})^{T}  ,\quad(\epsilon = \pm) \label{eq:Cmat-defn} \\
  \Cmat_{\epsilon}  \gamma_{5} \Cmat_{\epsilon}^{-1} &= \eta_{5} (\gamma_5)^{T},
\end{align}
for the Clifford algebra in even spacetime dimensions $D$.
These matrices can be defined in any even dimension, and we refer the reader to Ref.~\cite{stone_gamma_2021} for a detailed pedagogical discussion.\footnote{In Ref.~\cite{stone_gamma_2021} the $C_{+}, C_{-}$ matrices are referred to as $T$ and $C$ matrices, respectively.}
The action of $C_{\epsilon}$ on the ``fifth'' gamma matrix depends on the dimension, with $\eta_{5} = (-1)^{\frac{D}{2}}$.
With this definition, a charge conjugation symmetry $\Csym$ acts on a Dirac field operator $\psi(x)$ as
\begin{align}
	\Csym_{\epsilon}:
  \begin{cases}
  \psi \to \Cmat_{\epsilon} \overline{\psi}^{T} = \Cmat_\epsilon (\gamma^0)^T \psi^*\\
  \bar\psi \to \epsilon \psi^T \Cmat_{\epsilon}^{-1}
  \end{cases}
  \label{eq:Csym-defn}
\end{align}
where we used \cref{eq:Cmat-defn} to obtain the action on $\bar\psi$.
We can check that either choice of $\epsilon = \pm$ leads to a symmetry of the fermion kinetic term, since it transforms as:
\begin{align}
	\overline{\psi} \gamma^{\mu} \del_{\mu} \psi
  &\xrightarrow{\ \Csym_{\epsilon}\ } \epsilon \overline{\psi} (C_{\epsilon}^{-{1}}\gamma^\mu C_{\epsilon})^{T} \del_\mu \psi =  \overline{\psi} \gamma^\mu \del_\mu \psi.
\end{align}
On the other hand, a fermion mass term may not be invariant under this symmetry.
A general fermion bilinear transforms as
\begin{align}
	\overline{\psi} \Gamma \psi \xrightarrow{\ \Csym_{\epsilon}\ } \overline{\psi} (- \epsilon C_\epsilon^{-1} \Gamma C_\epsilon)^{T} \psi.
\end{align}
The transformation of various mass bilinears $\overline{\psi} \Gamma \psi$ for the $\Cmat_{\epsilon}$ symmetry is given in \cref{tab:cpt}.

  We note that the $\Csym$-conjugation on a Weyl fermion flips its chirality in 3+1d but not in 1+1d.
  Let $\psi$ be a right-chiral ($\gamma_5 = +1$) Weyl fermion field.
  Right-chirality of $\psi$ implies that $\gamma_5 \psi = + \psi$, which upon charge-conjugation implies
  \begin{align}
    \psi_{c} &= C (\gamma^0)^T \gamma_5^* \psi^{*}
      = - \eta_5 \gamma_5 \psi_{c},
  \end{align}
  where we used $\gamma_5^T = \gamma_5^{*}$.
  Therefore, we find that $\gamma_5\psi_{c} = - \eta_5 \psi_{c}$.
  In 3+1d with $\eta_5 = 1$, chirality flips under $\Csym$.
  However, this is different in 1+1d with $\eta_5 = -1$:
  the $\Csym$-conjugate of a left-chiral ($\gamma_{5}=-1$), right-moving Weyl fermion in $1+1$d is a left-chiral, right-moving Weyl antifermion.

\begin{table}
\setlength{\tabcolsep}{8pt}
  \begin{tabular}{ c | c c c c c   }
    \TopRule
    $\Gamma$ & $1$ & $\gamma^{5} $ & $\gamma^{\mu}$ & $\gamma^{\mu} \gamma^{5}$\\
    \MidRule
    $\Csym_{\epsilon}$ & $-\epsilon$ & $- \epsilon \eta_{5}$ & $-1$ & $\eta_{5}$\\
    $\Psym$ & 1 & $-1$ & $(-1)^{\mu}$ & $-(-1)^{\mu} $\\
    \BotRule
  \end{tabular}
  \caption{Charge-conjugation $\Csym_{\epsilon}$ and parity $\Psym$ transformation properties of various fermion bilinears $\overline{\psi} \Gamma \psi$. The two definitions of charge conjugation correspond to $\epsilon = \pm$, as defined in \cref{eq:Cmat-defn,eq:Csym-defn}. The action of charge-conjugation on $\gamma_{5}$ depends on the dimension and is given by $\eta_{5} = (-1)^{\frac{D}{2}}$ in $D$ spacetime dimensions. We use the notation where $(-1)^{\mu} = 1$ for $\mu =0 $ and $-1$ otherwise.}
  \label{tab:cpt}
\end{table}

\subsection{Lattice}

\minisec{Charge conjugation}
In 1+1 dimensions (mostly minus signature) with the basis choice $\gamma_{0}, \gamma_{1}, \gamma_{5} = \sigma_{3}, i \sigma_{2}, \sigma_{1} $, we obtain the $C_{\pm}$ matrices,
\begin{align}
	C_+ = \sigma_{3}, \quad C_{-} = \sigma_{2},
\end{align}
which lead to the following action of charge-conjugation $\Csym_{\pm}$ on the continuum Dirac fermion fields:
\begin{align}
	\Csym_{+}: \quad \psi &\to C_+ \gamma_0^T \psi^{*} = \psi^{*} \nonumber\\
	\Csym_{-}: \quad \psi &\to C_- \gamma_0^T \psi^{*} = i \sigma_{1} \psi^{*}.
                      \label{eq:Cpm}
\end{align}
In the staggered fermion formulation, the action in \cref{eq:Cpm} implies that $\Csym_{+}$ is an onsite symmetry, while $\Csym_{-}$ mixes with translations:
\begin{align}
	\Csym_{+}:&\quad \chi_n \to {\chi}^{\dagger}_{n}\label{eq:charge_conj_+} \\
	\Csym_{-}:&\quad \chi_n \to i {\chi}^{\dagger}_{n+1}
\end{align}
One can check that, with the above definitions, the transformations for $\Csym_{\pm}$ in \cref{tab:cpt} are satisfied on the 1+1-dimensional lattice for our discretization of the complex mass term as in \cref{eq:hpos}.


%% file: app_observables.tex
\section{Phase asymmetry for Gaussian wavepackets}
\label[appendix]{sec:observables}

In this appendix, we show how the phase asymmetry can be computed from the charge densities under the assumption of Gaussian wavepackets.

\minisec{Phase asymmetry in the continuum}
We begin with the incident wavefunction at $t \to -\infty$,
\begin{align}
\varphi_0(x,t)= \int dk\, f(k; \kwp, \sigmakwp)\, e^{i k(x-t)},
\end{align}
where $f(k; k_0,\sigma_k)=\mathcal{N}(\sigma_k)\,e^{-(k-k_0)^2/(4\sigma_k^2)}$, with $\mathcal{N}(\sigma_k)\equiv \ab[(2\pi)^{3/4}\sqrt{\sigma_k}]^{-1}$, such that $\int dk\,|f(k)|^2 = 1/(2\pi)$. 
With the reflection coefficient defined as $R(k)\equiv e^{-r(k)} e^{-i\phi(k)}$,  the reflected wavefunction as $t \to \infty$ can be written as
\begin{align}
    \varphi_r(x,t) &= \int dk\, R(k)\, f(k;k_0,\sigma_k)\,e^{-ik(x+t)}\nonumber\\
    &=\int dk\, \left[e^{-r(k)}\,f(k;k_0,\sigma_k)\right]\,e^{-ik(x+t)-i\phi(k)}\nonumber\\
    &=\mathcal{N}(\sigma_k)\int dk\, e^{-g(k)}\,e^{-ik(x+t)-i\phi(k)}.
    \label{eq:psirefl}
\end{align}
The peak of the momentum space wavefunction of the reflected wavepacket is determined by the minimum of the exponential factor $g(k)\equiv r(k)+(k-k_0)^2/(4\sigma_k^2)$ at $\overline{k_0}$, which satisfies the following equation:
\begin{align}
    0=g'(\kbar)=r'(\kbar) + \frac{\kbar-k_0}{2\sigma_k^2}.
\end{align}
To further simplify \cref{eq:psirefl}, we expand $g(k)$ and $\phi(k)$ around $\kbar$ and keep only the leading-order terms,
\begin{align}
    g(k)\approx g(\kbar) + \frac{1}{2}\left[r''(\kbar) + \frac{1}{2\sigma_k^2}\right](k-\kbar)^2,
\end{align}
\begin{align}
\phi(k)\approx \phi(\kbar) + \phi'(\kbar)(k-\kbar).
\end{align}
Substitute these expansions into \cref{eq:psirefl}, and define
\begin{align}
    &\frac{1}{4\sigmabar^2}\equiv \frac{1}{2}\left[r''(\kbar) + \frac{1}{2\sigma_k^2}\right],\\
    &\fbar(k;\kbar,\sigmabar) \equiv \mathcal{N}(\sigmabar)\,e^{-(k-\kbar)^2/(4\sigmabar^2)},
\end{align}
we get
\begin{align}
    \varphi_r(x,t) \approx e^{-g(\kbar)-i\zeta}\sqrt{\frac{\sigmabar}{\sigma_k}}\int dk\,\fbar(k;\kbar,\sigmabar)\, e^{-ik\left[x+t+\phi'(\kbar)\right]},
\end{align}
with $\zeta\equiv \phi(\kbar)-\phi'(\kbar)\kbar$.
The reflected particle number density after a sufficiently long time is
\begin{align}
    \rho_r(x,t) &= |\varphi_r(x,t)|^2 \\
    &=\sqrt{\frac{2}{\pi}}\,e^{-2g(\kbar)}\frac{\sigmabar^2}{\sigma_k}\, e^{-\left[x+t+\phi'(\kbar)\right]^2/(2\sigmaxbar^2)}\nonumber \\
    &= \frac{\sigmabar}{\sigma_k} e^{-2g(\kbar)}   N_{\sigmaxbar}(x + t; \phi'(\kbar))
      \label{eq:rho-r-gaussian}
    \end{align}
    with
\begin{align}
    \sigmaxbar\equiv\frac{1}{2\sigmabar},
\end{align}
and where we have used the notation $N_{\sigma}(x;\mu)$ to denote a normalized Gaussian with width $\sigma$ and center $\mu$.
\Cref{eq:rho-r-gaussian} shows that the reflected wave is an unnormalized Gaussian with amplitude reduced and the central position shifted to $\phi'(\bar k_{0})$.
For later use, we note that
\begin{align}
  \int dx\ \rho_{r}(x,t) &=
                     {\frac{\sigmabar}{\sigma_{k}} e^{-2 g(\kbar)} }
                     ,
                     \label{eq:Qtot}
  \\
 \int dx\, \rho_r(x,t)^{2} &= \frac{\sigmabar}{\sqrt{\pi}}{\ab[\int dx\, \rho_{r}(x,t) ]^{2}}.
                                                         \label{eq:sigmakbar}
\end{align}

In the case of charge symmetry breaking, the particle and antiparticle have different reflection amplitude $r_{\pm}(k)$ and phase $\phi_{\pm}(k)$. The difference in the reflection amplitude can be measured by the reflected charge in the symmetric phase, which, according to \cref{eq:Qtot}, is determined by $r_{\pm}(k)$.
To measure the difference in the phase, we calculate the product of the particle and antiparticle number densities and find it to be related to the phase difference:
\begin{align}
    -\int_{-\infty}^{0} dx\,\rho_+\,\rho_-= A\, e^{-B\asymPf^2}
\end{align}
where
\begin{align}
\asymPf &\equiv \phi'(\kbar_+) - \phi'(\kbar_-), \nonumber\\
    A  &= \sqrt{\frac{2}{\pi}}\frac{\sigmabar_+^2 \sigmabar_-^2}{\sigma_k^2\sqrt{\sigmabar_+^2 + \sigmabar_-^2}}\, e^{-2\left[g(\kbar_+) + g(\kbar_-)\right]},\nonumber\\
    B  &= \frac{2\sigmabar_+^2\sigmabar_-^2}{\sigmabar_+^2+\sigmabar_-^2}.
         \label{eq:asymP-AB}
\end{align}
From this we can define a new observable for the phase shift as
\begin{align}
  \asymPf
  &= \sqrt{- \frac{1}{B} \ln\left[ - \frac{\int_{-\infty}^{0} dx\  \rho_+\, \rho_{-} }{A}\right]}.
            \label{eq:dphi}
\end{align}

\minisec{Phase asymmetry on the lattice}
On the lattice, $\sigmabar_{\pm}$ and $e^{-2g(\kbar_{\pm})}$ can be measured from the discrete version of \cref{eq:Qtot} and \cref{eq:sigmakbar} as follows:
\begin{align}
  \sigmabar_{\pm} &= \frac{\sqrt{\pi}}{a} \frac{\sum_n (\rho_{\pm}^n)^2}{\ab(\sum_{n}\rho_{\pm}^n)^2},
    \nonumber \\
    e^{-2g(\kbar_{\pm})} &= \frac{\sigma_k }{\sigmabar_{\pm}} \sum_{n}\rho_{\pm}^n.
    \label{eq:asymP-lat-sigma}
\end{align}
With these quantities, one can calculate the coefficients $A$ and $B$. The cross term can be measured as
\begin{align}
  \int_{-\infty}^{0} dx\,\rho_+\rho_- \rightarrow
  \frac{1}{a} \sum_{n}w_n'\,\rho_+^n \rho_-^n,
    \label{eq:asymP-lat}
\end{align}
with the weight function $w_n'$ picking up only the charge density in the symmetric phase. This can then be used to calculate the phase shift observable using \cref{eq:dphi}.

\minisec{Some useful identities}
The product of two normalized Gaussians is an (unnormalized) Gaussian:
\begin{align}
  N_{\sigma_{1}}(x; \mu_{1}) N_{\sigma_{2}}(x; \mu_{2})
  &= N_{ \frac{{\sigma_1 \sigma_2}}{\overline{\sigma}}}(\mu_1 - \mu_2; 0) N_{\overline{\sigma}}(x; \overline{\mu})
\end{align}
with
\begin{align}
  \overline{\sigma} &= \sqrt{ \frac{\sigma_{1}^2 \sigma_2^2}{\sigma_1^2 + \sigma_2^{2}} }, \quad\quad
                 \overline{\mu} = \frac{\mu_1 \sigma_1^2 + \mu_2 \sigma_2^{2}}{\sigma_1^2 + \sigma_2^{2}}
\end{align}
In particular, to compute the square of a Gaussian, we can set $\sigma_1 = \sigma_2 = \sigma$ and $\mu_1 = \mu_{2} = \mu$ to get
\begin{align}
  N_{\sigma}(x; \mu)^2 &= N_{\sigma\sqrt{2}} (0; 0) N_{\frac{\sigma}{\sqrt{2}}}(x; \mu) = \frac{1}{2 \sigma \sqrt{\pi}} N_{\frac{\sigma}{\sqrt{2}}}(x; \mu)
\end{align}


%% file: app_wavepacket.tex
\section{Wavepackets on the lattice and time-evolution}
\label[appendix]{sec:create-wp}
We prepare the initial wavepacket as a superposition of plane waves defined on the full lattice in the absence of the bubble wall.
Taking \ac{PBC}, the Fourier transform takes the following form on the lattice for the staggered fermion:
\begin{align}
\zeta_k=
    \begin{bmatrix}
        \zeta_1 \\ \zeta_2 
    \end{bmatrix}_k \equiv \frac{1}{\sqrt{N/2}}\sum_{j=1}^{N/2}\begin{bmatrix}
        \chi_{2j-1}\,e^{-i(2j-1)k} \\ \chi_{2j}\,e^{-i 2jk}
    \end{bmatrix},
    \label{eq:kmodelat}
\end{align}
The momentum modes satisfying \ac{PBC} take the discrete values $k=-\frac{\pi}{2}+\frac{2\pi j}{N}$, $j=1,2,...,N/2$.
This leads to the lattice Hamiltonian given by:
\begin{align}
    \hat{H}_{\rm PBC} = \sum_{k}\zeta_k^\dagger\begin{pmatrix}
        0 & \sin k \\
        \sin k & 0
    \end{pmatrix}\zeta_k.
\end{align}
We define $(c_{k}^{\mathstrut}, d_{-k}^{\dagger})^T\equiv V_k\zeta_k^T$, with $V_k$ given by
\begin{align}
    V_k=\frac{1}{\sqrt{2}}\begin{pmatrix}
        1 & {\rm sign}(k) \\
        -{\rm sign}(k) & 1
    \end{pmatrix},
    \label{eq:def_Vk}
\end{align}
the Hamiltonian can be diagonalized as:
\begin{align}
    \hat{H}_{\rm PBC} &= \sum_k\epsilon_k\left(c_k^\dagger c_{k}^{\mathstrut} - d_{-k}^{\mathstrut} d_{-k}^{\dagger}\right)\nonumber\\&=\sum_k\epsilon_k\left(c_k^\dagger c_{k}^{\mathstrut} + d_{-k}^{\dagger} d_{-k}^{\mathstrut}\right)+{\rm const},\label{eq:diagH}
\end{align}
with $\epsilon_k=|\sin k|$.
$c_k^\dagger$ and $d_k^\dagger$ can be expressed in terms of the position space operators as,
\begin{align}
  c_{k}^\dagger&=\frac{1}{\sqrt{N}}
    \sum_{n=1}^{N}e^{ik n}\left[\Pi_{n0}+{\rm sign}(k)\Pi_{n1}\right]\chi_n^\dagger,
    \label{eq:ck}\\
  d_{k}^\dagger&=\frac{1}{\sqrt{N}}
    \sum_{n=1}^{N}e^{ik n}\left[\Pi_{n1}+{\rm sign}(k)\Pi_{n0}\right]\chi_n,
    \label{eq:dk}
\end{align}
where
\begin{align}
\Pi_{nl}\equiv\frac{1-(-1)^{n+l}}{2},~~l\in\{0,1\}.    
\end{align}
When acting on the vacuum, $\psiLat^\dagger_{k+}\equiv c_k^\dagger$ and $\psiLat^\dagger_{k-}\equiv d_k^\dagger$ create particle and antiparticle excitations with $P^\mu=(\epsilon_k, k)$.
We can then construct the creation operators for the fermion ($C^\dagger$) and antifermion ($D^\dagger$) wavepackets:
\begin{align}
    &C^\dagger=\sum_k \psiLat^\dagger_{k+}\,f^{\mathstrut}_{k} =\sum_n \chi_{n}^\dagger\,\varphi^{\mathstrut}_{n+},\label{eq:Cdagger}\\
    &D^\dagger=\sum_k \psiLat^\dagger_{k-}\, f^{\mathstrut}_{k}=\sum_n \chi_{n}\,\varphi^{\mathstrut}_{n-},
    \label{eq:Ddagger}
\end{align}
with
\begin{align}
    \varphi^{\mathstrut}_{n+}&=\frac{1}{\sqrt{N}}\sum_k f^{\mathstrut}_{k} e^{ikn}\left[\Pi_{n0}+{\rm sign}(k)\Pi_{n1}\right],\\
    \varphi^{\mathstrut}_{n-}&=\frac{1}{\sqrt{N}}\sum_k f^{\mathstrut}_{k} e^{ikn}\left[\Pi_{n1}+{\rm sign}(k)\Pi_{n0}\right].
\end{align}
The initial wavepacket $\ket|\Psi_{\pm}(0)>$ is thus prepared by applying \cref{eq:Cdagger} or \cref{eq:Ddagger} to the ground state of the Hamiltonian in \cref{eq:hpos}.

\minisec{Time evolution}
Staggered fermions in 1+1-dimensions can be mapped to a 1-dimensional spin chain by Jordan-Wigner transformation \cite{Sachdev_2011},
\begin{equation}
\chi_n=\left(\prod_{s<n}i\sigma^z_s\right)\sigma_n^+,~~~~~\chi_n^\dagger=\left(\prod_{s<n}-i\sigma^z_s\right)\sigma_n^-,\\
\end{equation}
which ensures anti-commutation relations between different lattice sites.
This leads to the following Hamiltonian:
\begin{align}
 \hat{H}=& \sum^{N-1}_{n=1} h_{n,n+1} + h_N,
  \label{eq:jw_h}
 \end{align}
with $h_N =(-1)^{N+1}|m_N^{\mathstrut}|\cos\theta^{\mathstrut}_{N}\sigma_{N}^{-}\sigma_{N}^{+}$ and
\begin{align}
    h_{n,n+1}=&\left[\frac{1}{2}+(-1)^n |m_n|\sin\theta_n\right](\sigma_{n+1}^-\sigma_{n}^+ + \sigma_n^- \sigma_{n+1}^+)\nonumber\\&-(-1)^n |m_n|\cos\theta^{\mathstrut}_{n}\sigma_n^- \sigma_n^+.
    \label{eq:hdim}
    \end{align}
The time evolution of this Hamiltonian can be efficiently implemented by Trotter decomposition. For our simulation, we employ second-order trotterization \cite{Hatano:2005gh}:
\begin{align}
    e^{-i\hat{H}\tauh}&\approx e^{-ih_{1,2}^{\mathstrut}\tauh/2}\,e^{-ih_{2,3}^{\mathstrut}\tauh/2}\cdots e^{-ih_{N-1,N}^{\mathstrut}\tauh/2}\,e^{-ih_N^{\mathstrut}\tauh}\nonumber\\
    &e^{-ih_{N-1,N}^{\mathstrut}\tauh/2}\cdots e^{-ih_{2,3}^{\mathstrut}\tauh/2}\,e^{-ih_{1,2}^{\mathstrut}\tauh/2} +\mathcal{O}(\tauh^3),
    \label{eq:trotter_op}
\end{align}
where $\tauh=\tau/a$ is the trotter step in lattice units.
The error for each time step is $\mathcal{O}(N\tauh^3)$ for trotterizing $H$ into $N$ pairs of non-commuting terms. For total evolution time $T$, there are $N_t = T/\tau$ trotter steps, thus the total error is $\mathcal{O}(N N_t \tauh^3)$.

\minisec{A note on boundary conditions and momentun truncation}
To ensure that the wavepacket moves towards the bubble wall and avoids scattering with the lattice boundary in the direction opposite to the bubble, we truncate away the non-positive $k$-modes from the wavepacket.
This truncation will distort the wavepacket from a perfect Gaussian shape, which will cause errors in measuring local observables.
For $\theta_0 = 0$, where we can compare with exact analytic results, we find this truncation errors to be negligible.
When $\theta_{0} \neq 0$, where we study the asymmetry generation, we choose $k_0\geq 2\sigma_k$ to ensure that the negative modes lie beyond $2\sigma_k$ from $k_0$, thereby diminishing the Gaussian shape distortion.


%% file: app_systematics.tex
\section{Systematic Uncertainties}
\label[appendix]{sec:systematics}

In the simulations presented in \cref{fig:real-mass}--\ref{fig:phase-asym}, we fix the physical volume to $\Lphys=28$ and extrapolate to the continuum limit $a \to 0$.
In this appendix, we investigate the systematic uncertainties associated with the finite volume effects, which could be a dominant source of errors.
We vary lattice spacing in the range
$a\in\{\frac{1}{8},\frac{1}{6},\frac{1}{5},\frac{1}{4}\}$, and physical volume $\Lphys \in \{20, 24, 28\}$.

\subsection{Charge asymmetry of the initial state}
\label{sec:errors-Csym-initial}

\begin{figure*}
    \centering
    \includegraphics[width=0.95\linewidth]{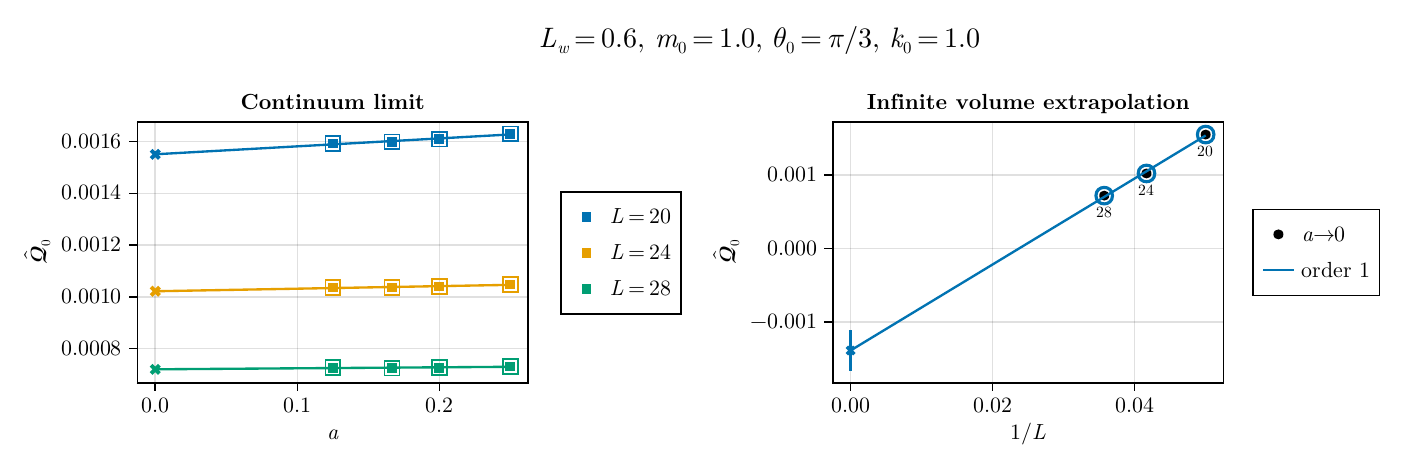}
    \caption{\emph{Magnitude asymmetry} of the initial state $\asymQL{0}$ for various lattice spacings and physical volumes. The mass profile is chosen to be the same one used for \cref{fig:mag-asym} and \cref{fig:phase-asym}, with $\theta_0=\pi/3$. For each physical volume $L$, we calculate the continuum limits by fitting $\asymQL{0}$ as a linear function of the lattice spacings $a$, and extrapolating to $a=0$. The continuum limits are then fitted as a linear function of the inverse lattice volume $1/L$ and extrapolated to the infinite volume limit $1/L\to 0$. The deviation of the infinite volume limit of $\asymQL{0}$ from zero measures the systematic uncertainty of the charge asymmetry in the initial state.}
    \label{fig:qinit_con_inf}
\end{figure*}

\begin{figure*}
    \includegraphics[width=0.95\linewidth]{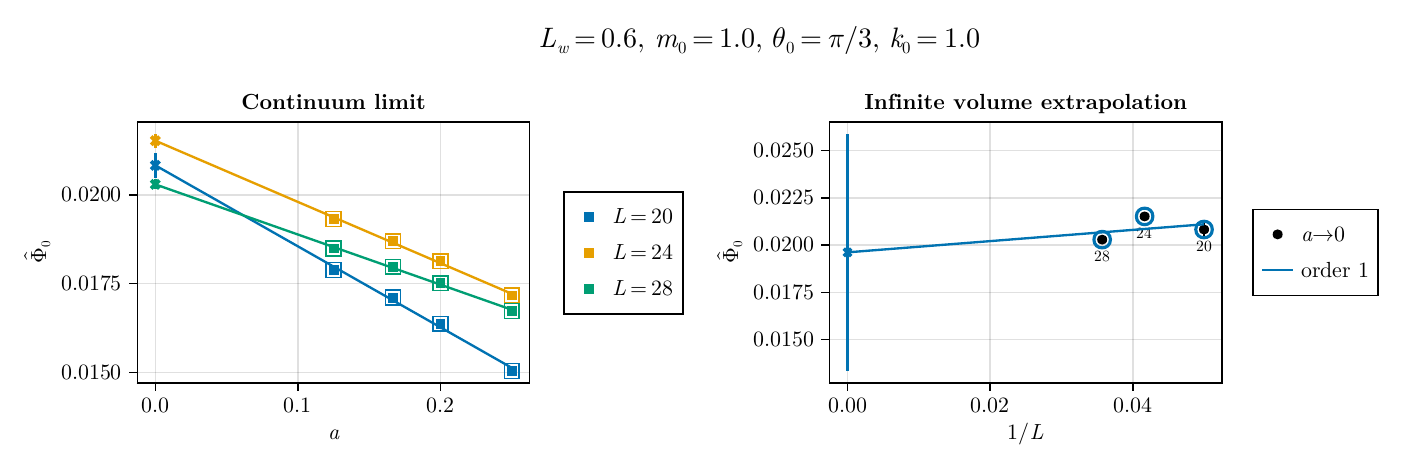}
    \caption{Same as \cref{fig:qinit_con_inf} but for the \emph{Phase asymmetry} $\asymPL{0}$ of the initial state.}
    \label{fig:phiinit_con_inf}
\end{figure*}

We first check the charge asymmetry of the initial state consisting of the particle and antiparticle wavepackets by measuring the magnitude asymmetry and phase asymmetry in the symmetric phase at $t=0$, $\asymQL{0}$ and $\asymPL{0}$.
Using the same hyperbolic complex mass profile as in \cref{fig:mag-asym} and \cref{fig:phase-asym}, we prepare the initial state and measure $\asymQL{0}$ and $\asymPL{0}$.
Since the wavepackets have not yet interacted with the bubble wall, the $\theta_0$ dependence of the initial state is expected to be very weak.
In \cref{fig:qinit_con_inf} and \cref{fig:phiinit_con_inf}, we show as an example for $\theta_0=\pi/3$, the measured values of $\asymQL{0}$ and $\asymPL{0}$ at various lattice spacings and physical volumes.
With these results, we first perform a linear fit of the two observables as a function of the lattice spacings $a$, and extrapolate to the continuum limit $a\to 0$, as shown in the left panels of \cref{fig:qinit_con_inf} and \cref{fig:phiinit_con_inf}. We then perform a linear fit of the continuum limits as a function of the inverse physical volumes $1/L$ and extrapolate to the infinite volume limit $1/L\to 0$, as shown in the right panels of \cref{fig:qinit_con_inf} and \cref{fig:phiinit_con_inf}. The extrapolation results are the ``$\mathbf{\times}$'' points in the plots, with the error bars showing $1\sigma$ fitting errors.
The deviation of the extrapolation results from zero quantifies the systematic uncertainties in the initial state preparation.

\subsection{Asymmetry generation from scattering with step function complex mass profile}

We also investigate the systematic uncertainties associated with the scattering.
For this investigation, we take the mass profile to be a step function for both magnitude and phase:
\begin{align}
|m(x)|= \begin{cases} 0& x\leq 0\\ m_0 & x > 0 \end{cases}~~,~~
\theta(x) = \begin{cases} 0 & x\leq 0\\ \theta_0 & x > 0 \end{cases}~~.
\end{align}
The complex phase of this profile, although nonvanishing in the broken phase, can be rotated away by a field redefinition without breaking other symmetries. Therefore, no asymmetry is expected to be generated in this case.
Any deviation from zero for the asymmetry observables $\asymQL{\infty}$ and $\asymPLf$ is thus a measure of systematic errors.
In the following, we show the results for step-function mass profile with $m_0=1$ and $\theta_0$ varying in the range $[0,\pi/2]$.
With the results for various $a$ and $L$, we first extrapolate to the continuum $a\to 0$ for each fixed $\Lphys$.
Then to study finite volume effects, we perform an infinite-volume extrapolation using a linear fit.

\begin{figure*}
    \centering
    \includegraphics[width=0.95\linewidth]{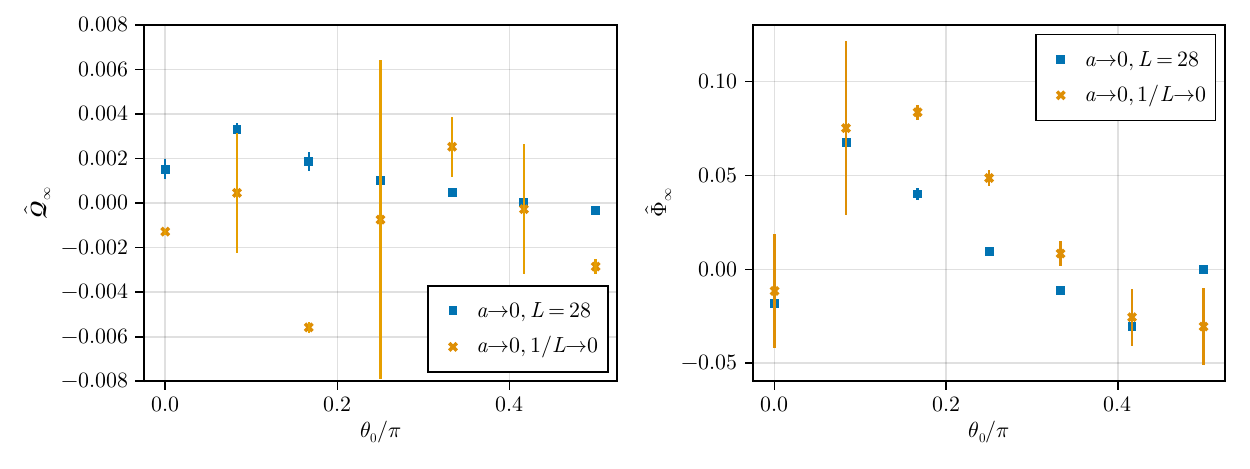}
    \caption{The continuum limits at the largest physical volume $L=28$, and the infinite volume limits of $\asymQLf$ and $\asymPLf$ generated from scattering with the step mass profiles with $m_0=1$ and $\theta_0$ varying in the range $[0,\pi/2]$. The error bars represent the $1\sigma$ fitting errors.}
    \label{fig:Qs-dphi-step-thetavary}
\end{figure*}

In \cref{fig:Qs-dphi-step-thetavary}, we show the continuum limits at the largest physical volume $\Lphys=28$, as well as the infinite volume extrapolations of $\asymQLf$ and $\asymPLf$, where the error bars represent $1\sigma$ errors from the infinite-volume linear fits.
As expected from a step-function profile, the absolute values of $\asymQLf$ and $\asymPLf$ are close to zero for $\Lphys = 28$.
The measured magntitude asymmetry is of the order $\asymQLf \sim 10^{-3}$ while the phase asymmetry is $\asymPLf \sim 10^{-2}$. This therefore provides an approximate measure of the absolute errors in our studies of the complex mass profile in \cref{sec:results-complex}.
We expect that the nonvanishing asymmetry is likely due to finite-volume effects.
Looking at the infinite-volume results in \cref{fig:Qs-dphi-step-thetavary}, we observe that a simple linear $1 / \Lphys \to 0$ extrapolation is not under control, sometimes leading to larger uncertainties compared to the continuum limit results at $\Lphys=28$.
Therefore, in this work, we work with a fixed physical volume of $\Lphys=28$ and leave a more systematic analysis of the finite-volume effects for future work.

